\documentclass[10pt,journal]{IEEEtran}

\usepackage{amsmath,amsfonts}
\usepackage{algorithmic}
\usepackage{algorithm}
\usepackage{array}
\usepackage[caption=false,font=normalsize,labelfont=sf,textfont=sf]{subfig}
\usepackage{textcomp}
\usepackage{stfloats}
\usepackage{url}
\usepackage{verbatim}
\usepackage{graphicx}
\usepackage{subfig}
\usepackage{stmaryrd}

\usepackage{tabularray}

\usepackage{bm} 

\usepackage{acro} 

\RenewAcroTemplate[list]{description}{%
	\acroheading
	\acropreamble
	\begin{LaTeXdescription}
		\acronymsmapF{%
			\item[\acrowrite{short}\acroifT{alt}{/\acrowrite{alt}}]
			\acrowrite{list}%
			\acroifanyT{foreign,extra}{ (}%
			\acroifT{foreign}{\acrowrite{foreign}\acroifT{extra}{, }}%
			\acroifT{extra}{\acrowrite{extra}}%
			\acroifanyT{foreign,extra}{)}%
			\acropagefill
			\acropages
			{\acrotranslate{page}\nobreakspace}
			{\acrotranslate{pages}\nobreakspace}%
		}
		{\item\AcroRerun}
	\end{LaTeXdescription}
}
\usepackage[
	style=ieee,
	dashed=false,
	maxbibnames=20,
	minbibnames=2,
	maxcitenames=1,
	mincitenames=1,
	backend=biber,
	sorting=none,
	useprefix=false
	]{biblatex}

\usepackage{tikz}
\usetikzlibrary{arrows,shapes,intersections,calc,angles,decorations.pathreplacing,intersections,quotes,angles,positioning,fit,petri,through,shadows,plotmarks,spy}
\newlength{\unit}

\pdfoutput=1

\DeclareAcronym{2D}{
	short=2-D,
	long=2-dimensional,
}

\DeclareAcronym{3D}{
	short=3-D,
	long=3-dimensional,
}

\DeclareAcronym{4D}{
	short=4-D,
	long=4-dimensional,
}

\DeclareAcronym{nD}{
	short=$n$-D,
	long=$n$-dimensional,
}

\DeclareAcronym{CT}{
	short=CT, 
	long=computed tomography,
}

\DeclareAcronym{PET}{
	short=PET, 
	long=positron emission tomography,
}

\DeclareAcronym{MRI}{
	short=MRI, 
	long=magnetic resonance imaging,
}

\DeclareAcronym{MR}{
	short=MR, 
	long=magnetic resonance,
}

\DeclareAcronym{HR}{
	short=HR, 
	long=high-resolution,
}

\DeclareAcronym{PCCT}{
	short=PCCT, 
	long=photon-counting computed tomography,
}

\DeclareAcronym{DECT}{
	short=DECT, 
	long=dual-energy computed tomography,
}

\DeclareAcronym{PCD}{
	short=PCD, 
	long=photon-counting detectors,
}

\DeclareAcronym{EID}{
	short=EID, 
	long= energy-integrating detectors,
}

\DeclareAcronym{FOV}{
	short=FOV, 
	long=field of view,
}

\DeclareAcronym{FBP}{
	short=FBP, 
	long=filtered backprojection,
}

\DeclareAcronym{MBIR}{
	short=MBIR, 
	long=model-based iterative reconstruction,
}

\DeclareAcronym{MLAA}{
	short=MLAA, 
	long=maximum-likelihood attenuation and activity,
}

\DeclareAcronym{BM3D}{
	short=BM3D, 
	long=block-matching and 3-D filtering,
}

\DeclareAcronym{WLS}{
	short=WLS, 
	long=weighted least squares,
}
\DeclareAcronym{PWLS}{
	short=PWLS, 
	long=penalized weighted least squares,
}

\DeclareAcronym{MSE}{
	short=MSE, 
	long=mean squared error,
}

\DeclareAcronym{MAE}{
	short=MAE, 
	long=mean absolute error,
}

\DeclareAcronym{CNR}{
	short=CNR, 
	long=contrast-to-noise ratio,
}

\DeclareAcronym{CS}{
	short=CS,
	long=compressed sensing,
}

\DeclareAcronym{TV}{
	short=TV,
	long=total variation,
}

\DeclareAcronym{LR}{
	short=LR,
	long=low-rank,
}

\DeclareAcronym{TNV}{
	short=TNV, 
	long=total nuclear variation,
}

\DeclareAcronym{JTV}{
	short=JTV, 
	long=joint total variation,
}

\DeclareAcronym{DTV}{
	short=DTV, 
	long=directional total variation,
}

\DeclareAcronym{PLS}{
	short=PLS, 
	long=parallel level sets,
}

\DeclareAcronym{SQS}{
	short=SQS, 
	long=separable quadratic surrogate,
}

\DeclareAcronym{ADMM}{
	short=ADMM, 
	long=alternating direction method of multipliers,
}

\DeclareAcronym{ISTA}{
	short=ISTA, 
	long=iterative soft thresholding algorithm,
}

\DeclareAcronym{SVT}{
	short=SVT, 
	long=singular value thresholding,
}

\DeclareAcronym{DL}{
	short=DL,
	long=dictionary learning,
}

\DeclareAcronym{TDL}{
	short=TDL,
	long=tensor dictionary learning,
}

\DeclareAcronym{CDL}{
	short=CDL,
	long=convolutional dictionary learning,
}

\DeclareAcronym{MCDL}{
	short=CDL,
	long=multichannel convolutional dictionary learning,
}

\DeclareAcronym{CAOL}{
	short=CAOL, 
	long=convolutional analysis operator learning,
}

\DeclareAcronym{MCAOL}{
	short=MCAOL, 
	long=multichannel convolutional analysis operator learning,
}

\DeclareAcronym{AI}{
	short=AI, 
	long=artificial intelligence,
}
\DeclareAcronym{AGI}{
	short=AGI, 
	long=artificial general intelligence,
}

\DeclareAcronym{DIP}{
	short=DIP, 
	long=deep image prior,
}

\DeclareAcronym{DLIR}{
	short=DLIR, 
	long=deep learning image reconstruction,
}

\DeclareAcronym{NN}{
	short=NN, 
	long=neural network,
}

\DeclareAcronym{CNN}{
	short=CNN, 
	long=convolutional neural network,
}

\DeclareAcronym{RNN}{
	short=RNN, 
	long=recurrent neural network,
}
\DeclareAcronym{FCN}{
	short=FCN,
        long=fully convolutional network,
}

\DeclareAcronym{GAN}{
	short=GAN, 
	long=generative adversarial network,
}

\DeclareAcronym{WGAN}{
	short=W-GAN, 
	long=Wasserstein generative adversarial network
}

\DeclareAcronym{AE}{
	short=AE, 
	long=auto-encoder,
}

\DeclareAcronym{SAE}{
	short=SAE,
	long=stacked auto-encoder,
}

\DeclareAcronym{VAE}{
	short=VAE, 
	long=variational auto-encoder,
}

\DeclareAcronym{CED}{
	short=CED, 
	long=convolutional encoder-decoder, 
}

\DeclareAcronym{LLE}{
	short=LLE, 
	long=locally linear embedding ,
}

\DeclareAcronym{PRISM}{
	short=PRISM, 
	long=prior rank intensity and sparsity model,
}

\DeclareAcronym{PICCS}{
	short=PICCS, 
	long=prior image-constrained compressed sensing ,
}

\DeclareAcronym{CPD}{
	short=CPD, 
	long=canonical polyadic decomposition,
}

\DeclareAcronym{VMI}{
	short=VMI, 
	long=virtual monochromatic image,
}

\DeclareAcronym{mGy}{
	short=mGy, 
	long=milligray ,
}

\DeclareAcronym{keV}{
	short=keV, 
	long=kilo electronvolt,
}

\DeclareAcronym{MeV}{
	short=MeV, 
	long=mega electronvolt,
}

\DeclareAcronym{kVp}{
	short=kVp, 
	long=peak kilovoltage,
}

\DeclareAcronym{SNR}{
	short=SNR, 
	long=signal-to-noise ratio,
}

\DeclareAcronym{HU}{
	short=HU, 
	long=Hounsfield unit,
}

\DeclareAcronym{FDA}{
	short=FDA, 
	long=Food and Drug Administration,
}

\DeclareAcronym{DM}{
	short=DM, 
	long=diffusion model,
}

\DeclareAcronym{CZT}{
	short=CZT, 
	long=cadmium zinc telluride,
}

\renewcommand{\hat}{\widehat}
\renewcommand{\bar}{\overline}

\newcommand{\bolda}{\bm{a}}
\newcommand{\boldb}{\bm{b}}

\newcommand{\boldd}{\bm{d}}

\newcommand{\boldg}{\bm{g}}

\newcommand{\boldr}{\bm{r}}

\newcommand{\boldx}{\bm{x}}
\newcommand{\boldy}{\bm{y}}
\newcommand{\boldz}{\bm{z}}

\newcommand{\boldmu}{\bm{\mu}}

\newcommand{\boldA}{\bm{A}}

\newcommand{\boldD}{\bm{D}}
\newcommand{\boldE}{\bm{E}}
\newcommand{\boldF}{\bm{F}}

\newcommand{\boldH}{\bm{H}}

\newcommand{\boldM}{\bm{M}}

\newcommand{\boldP}{\bm{P}}

\newcommand{\boldT}{\bm{T}}

\newcommand{\calD}{\mathcal{D}}

\newcommand{\calL}{\mathcal{L}}
\newcommand{\calM}{\mathcal{M}}

\newcommand{\calP}{\mathcal{P}}

\newcommand{\calZ}{\mathcal{Z}}

\newcommand{\bbE}{\mathbb{E}}

\newcommand{\boldcalF}{\bm{\mathcal{F}}}
\newcommand{\boldcalG}{\bm{\mathcal{G}}}

\newcommand{\boldcalL}{\bm{\mathcal{L}}}

\newcommand{\boldcalR}{\bm{\mathcal{R}}}

\newcommand{\rmd}{\mathrm{d}}
\newcommand{\rme}{\mathrm{e}}

\newcommand{\rms}{\mathrm{s}}

\newcommand{\boldtheta}{\bm{\theta}}
\newcommand{\boldTheta}{\bm{\Theta}}

\newcommand{\boldphi}{\bm{\phi}}

\newcommand{\ybar}{\bar{y}}
\newcommand{\boldybar}{\bar{\boldy}}

\newcommand{\boldmuhat}{\hat{\boldmu}}
\newcommand{\boldxhat}{\hat{\boldx}}
\newcommand{\boldzhat}{\hat{\boldz}}

\renewcommand{\hbar}{\bar{h}}

\newcommand{\argmin}{\operatornamewithlimits{arg\,min}}

\newcommand{\eg}{e.g.,}

\newcommand{\transp}{^\top}

\newcommand{\R}{\mathbb{R}}

\renewcommand{\th}{\textsuperscript{th}}

\title{Systematic Review on Learning-based Spectral CT}

\author{Alexandre Bousse,~\IEEEmembership{Member,~IEEE}, Venkata Sai Sundar Kandarpa, Simon Rit, Alessandro Perelli, Mengzhou Li, Guobao Wang,~\IEEEmembership{Senior Member,~IEEE}, Jian Zhou,~\IEEEmembership{Senior Member,~IEEE}, Ge Wang,~\IEEEmembership{Fellow,~IEEE}
	
    \thanks{This work did not involve human subjects or animals in its research.}
	\thanks{This work was supported by the French National Research Agency (ANR) under grant No ANR-20-CE45-0020, and the National Institutes of Health (NIH) under grant Nos R01EB026646, R01CA233888, R01CA237267, R01HL151561, R42GM142394, R21CA264772, R01EB031102, R01EB032716, and R21EB027346.}
	\thanks{Alexandre Bousse and Venkata Sai Sundar Kandarpa are with Univ. Brest, LaTIM, Inserm, U1101, 29238~Brest, France.}
    \thanks{Simon Rit is with Univ. Lyon, INSA-Lyon, Universit{\'e} Claude Bernard Lyon 1, UJM-Saint {\'E}tienne, CNRS, Inserm, CREATIS UMR 5220, U1294, F-69373, Lyon, France.}
	\thanks{Alessandro Perelli is with the School of Science and Engineering, University of Dundee, DD1~4HN~Dundee, U.K.}
	\thanks{Guobao Wang is with the Department of Radiology, University of California Davis Health, Sacramento, CA~95817~USA.}
	\thanks{Jian Zhou is with the CTIQ, Canon Medical Research USA, Inc., Vernon Hills, IL~60061~USA}
	\thanks{Mengzhou Li and Ge Wang are with the Biomedical Imaging Center, Rensselaer Polytechnic Institute, Troy, NY~12180~USA.}
	\thanks{Corresponding authors: A. Bousse, \mbox{bousse@univ-brest.fr}} }

\begin{document}
	
\maketitle

\begin{abstract}
	Spectral \ac{CT} has recently emerged as an advanced version of medical \ac{CT} and significantly improves conventional (single-energy) \ac{CT}. Spectral \ac{CT} has two main forms: \ac{DECT} and \ac{PCCT}, which offer  image improvement, material decomposition, and feature quantification relative to conventional \ac{CT}. However, the inherent challenges of spectral \ac{CT}, evidenced by data and image artifacts, remain a bottleneck for clinical applications. To address these problems, machine learning techniques have been widely applied to spectral \ac{CT}. In this review, we present the  state-of-the-art data-driven techniques for spectral \ac{CT}. 
\end{abstract}


\printacronyms

\section{Introduction}

\IEEEPARstart{S}{ince} Cormack and Hounsfield's Nobel prize-winning breakthrough, X-ray \ac{CT} is extensively used in medical applications and produces a huge number of gray-scale \ac{CT} images. However, these images are often insufficient to distinguish crucial differences between biological tissues and contrast agents. From the perspective of physics, the X-ray spectrum from a medical device is polychromatic, and interactions between X-rays and biological tissues depend on the X-ray energy, which suggests the feasibility to obtain spectral, multi-energy, or true-color, \ac{CT} images.  

Over the past decade, spectral \ac{CT} has been rapidly developed as a new generation of \ac{CT} technology. \Ac{DECT} and \ac{PCCT} are the two main forms of spectral \ac{CT}. \Ac{DECT} is a method of acquiring two projection datasets at different energy levels. \Ac{PCCT}, on the other hand, uses detectors that measure individual photons and their energy, promising significantly better performance with major improvements in energy resolution, spatial resolution and dose efficiency \cite{taguchi2013vision,taguchi2021photon}. Despite the intrinsic merits of spectral \ac{CT}, there are technical challenges already, being or yet to be addressed \cite{taguchi2022model,tai2023effects}. To meet these challenges, the solutions can be hardware-oriented, software-oriented, or hybrid.

Traditionally, \ac{CT} algorithms are grouped into two categories, which are analytic and iterative reconstruction respectively. A new category of \ac{CT} algorithms has recently emerged: \ac{AI}-inspired, learning-based or data-driven reconstruction. These algorithms are commonly implemented as deep \acp{NN}, which are iteratively trained for image reconstruction and post-processing, and then used for inference in the feed-forward fashion just like a closed-form solution.

Several reviews have been dedicated to machine learning and deep learning in \ac{CT}. These papers cover a wide range of topics, including image reconstruction, segmentation, classification, and more. For example, \citeauthor{litjens2017survey}~\cite{litjens2017survey} and \citeauthor{sahiner2019deep}~\cite{sahiner2019deep} comprehensively surveyed deep learning applications in medical imaging. \citeauthor{domingues2020using}~\cite{domingues2020using} proposed a review on deep learning in \ac{CT} and \ac{PET}. However, few have specifically focused on spectral \ac{CT}. 

\begin{figure*}
	
	\centering
	\input{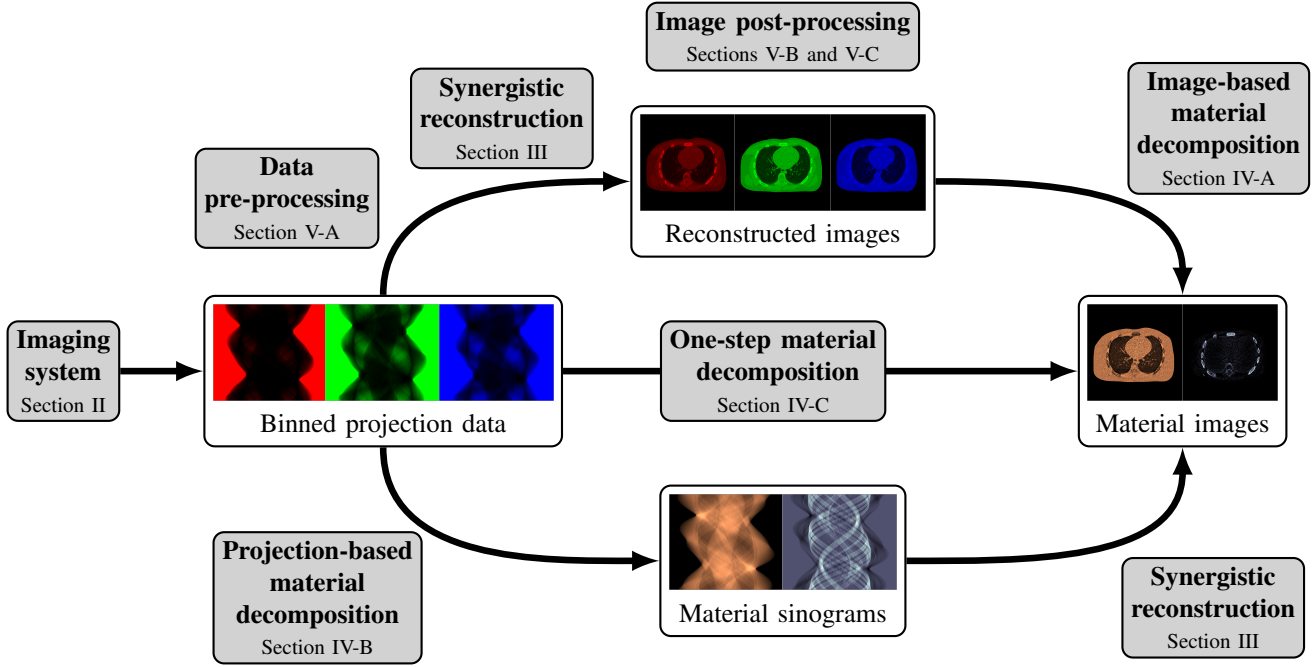}
	
	\caption{Structure of this review paper, with the sections keyed to the main steps in the spectral \ac{CT} imaging process. }\label{fig:tasks}
\end{figure*}

This review paper provides a technical overview of the current state-of-the-art of machine learning techniques for spectral \ac{CT}, especially deep learning ones. The paper is divided into the following sections:  
\ac{DECT} and \ac{PCCT} systems,
image reconstruction,
material decomposition,
pre- and post-processing,
hybrid imaging,
perspectives and conclusion. 
Section~\ref{sec:systems} describes \ac{DECT} and \ac{PCCT} systems. 
Section~\ref{sec:recon} discusses the application of learning-based techniques for multi-energy \ac{CT} reconstruction from energy-binned data, which use shallow or deep network architectures, from \ac{DL}  to much deeper contemporary network models. Reconstruction of multi-energy \ac{CT} images will face the problem of beam hardening.
Section~\ref{sec:mat_decomp} covers different approaches to material decomposition: image-based techniques, which use as input multi-energy \ac{CT}, and alternative solutions to beam hardening, projections-based and one-step decompositions.
Section~\ref{sec:im_proc} is dedicated to various pre-processing and post-processing aspects, which are based on sinogram data and spectral \ac{CT} images respectively, including data calibration, 
image denoising and  artifacts correction, as well as image generation.  
Finally, Section~\ref{sec:perspectives} covers key issues and future directions of learning-based spectral \ac{CT}. The structure  of this paper is outlined in Fig.~\ref{fig:tasks}.

\subsection*{Notations}

Vectors (resp. matrices) are represented with bold lowercase (resp. uppercase characters). Images are represented as  $J$-dimensional real-valued vectors which can be reshaped in \ac{2D} or \ac{3D} objects, where $J$ is the number of image pixels or voxels. $I$ is the number of rays per energy bin. `$\transp$' is the matrix transposition symbol. A \ac{NN} is represented by a bold calligraphic upper case character with a subscript representing the weights to be trained, e.g., $\boldcalF_{\boldtheta}$. $\|\cdot\|_0$ is the $\ell^0$ semi-norm defined for all $\boldx = [x_1,\dots,x_N]\transp\in\R^N$ as $\|\boldx\|_0 = \#\{n\in\{1,\dots,N\} \colon x_n \neq 0 \}$, where $\# A$ denotes the cardinal of set $A$, and $\|\cdot\|_p$, $p>1$ the $\ell^p$-norm. For a positive-definite matrix $\boldM\in\R^{N\times N}$, $\|\cdot\|_{\boldM}$ is the $\ell^2$ weighted-norm defined for all $\boldx \in\R^N$ as $ \|\boldx\|_{\boldM} =  \sqrt{ \boldx\transp \boldM \boldx  }   $, and $\|\cdot\|_{\mathrm{F}}$ denotes the Frobenius norm. $[\bolda , \boldb]$ is the horizontal concatenation of two column vectors $\bolda$ and $\boldb$ with the same length. $\{\boldx_k\} = \{\boldx_k, k = 1,\dots, K\}$  denotes an ordered collection of vectors where the number of elements $K$ depends on the context. $L(\cdot,\cdot)$ denotes a loss function that evaluates the adequation between 2 vectors, e.g, $L(\bolda,\boldb) = \sum_i -a_i \log b_i + b_i$ (negative Poisson log-likelihood), or $L(\bolda,\boldb) = \|\bolda - \boldb\|^p_p$. $R$ is a regularisation functional.

\section{DECT and PCCT Systems}\label{sec:systems}

The first attempt to differentiate materials using \ac{CT} with multiple X-ray energy spectra was made in the 1970s \cite{GENANT1977}. Since then, technologies in spectral \ac{CT}  have been continuously evolving.  Traditional \Ac{DECT} and spectrally-resolving \ac{PCCT} are the two specific  forms of spectral \ac{CT} that are both commercially available. The former uses a minimum of two separate X-ray energy spectra to differentiate two basis materials with different attenuation properties at various energy levels, while the latter usually involves the advanced detector technology known as energy resolving \ac{PCD}, which resolves spectral information of X-ray photons in two or more energy bins emitted from a polychromatic X-ray source. \Ac{DECT} overcomes several limitations of single energy spectrum \ac{CT} and has achieved clinical acceptance and widespread applications. In the following, several types of \ac{DECT} are briefly described. We will not cover all technologies, but we will focus on those that are currently representative. The interested readers may refer to~\cite{Adam2021-ap,Forghani2017-ra,Fornaro2011-cf,Johnson2012-il,Krauss2018-hv,McCollough2015-tx,Rajiah2017-pl,Johnson2010-pq,Megibow2018-xy,Greffier2022-yc} for more details and comparisons.

Sequential acquisition is perhaps the most straightforward \ac{DECT} imaging approach. It performs two consecutive or subsequent scans of the same anatomy using an X-ray source operated at a low-\ac{kVp} setting and then a high-\ac{kVp} setting. The approach requires no hardware modification, but may suffer from image mis-registration due to motion artifacts from the delay between low- and high-\ac{kVp} scans. Advanced \ac{DECT} technologies all utilize specific hardware to mitigate the misregistration problem and shorten the data acquisition time.

The dual-source \ac{DECT} scanner was first introduced in 2005~\cite{Flohr2006-ol}, which is featured by two source-detector systems orthogonally arranged in the same gantry to acquire the low- and the high-energy scan simultaneously. Although the 90-degree phase shift between the two scans creates a slight temporal offset, the two X-ray sources can select independent X-ray energy spectra to optimize the spectral separation for material differentiation in the data and/or image domains. 

A dual-layer detector or a combination of two detector layers of scintillation material is also a good solution for \ac{DECT}~\cite{Carmi05,Vlassenbroek2010,philips-dual-layer,Gabbai2015-wm}. In this approach, low- and high-energy datasets are collected simultaneously by the two detector layers with perfect spatial alignment and excellent synchronicity. This advantage simplifies direct data-domain material decomposition.

Fast \ac{kVp}-switching \ac{DECT} is yet another technology that uses a highly specialized X-ray generator that can rapidly switch the tube voltage between low- and high-\ac{kVp} settings during data acquisition. The first commercially available fast \ac{kVp}-switching \ac{DECT} scanner (GE~Discovery~CT750~HD) is capable of changing the tube voltage for each projection angle, so that each low- and high-\ac{kVp} projection can be obtained almost simultaneously. The material decomposition can then be performed in the data domain. A similar design has been reported in \cite{Li-MeV-DECT-2018} where the authors have utilized a linear accelerator as X-ray source to generate rapid switching electron pulses of 6~MeV and 9~MeV respectively. This has resulted in an experimental MeV~\ac{DECT} system that has been developed to perform cargo container inspection. Another type of fast \ac{kVp}-switching \ac{DECT} scanner has recently been introduced (Canon Aquilion ONE/PRISM)~\cite{Canon-kv-switching} that switches the tube voltage less frequently, allowing it to acquire the same energy from multiple successive projection angles. This design simplifies tube current modulation, making dose balancing at the two energy levels less complex. Along with the fast \ac{kVp}-switching process, there is also a grating-based method that can help improve data acquisition~\cite{Xi2017}. In this method, an X-ray filter that combines absorption and filtering gratings is placed between the source and the patient. The gratings move relative to each other and are synchronized with the tube switching process to avoid spectral correlation. Simulation studies have shown improved spectral information with reduced motion-induced artifacts. 

\ac{PCD} technology plays an important role in \ac{PCCT} imaging. \Acp{PCD} requires a single layer of semiconductor sensor that converts X-ray photons directly into electrical signals. The main converter materials at present are 
\ac{CZT} and Si. \ac{CZT} is a material with a higher atomic number Z than Si and has a relatively high X-ray stopping power. Thus, the \ac{CZT}-based \ac{PCD} can have thin sensor layers of only a few millimeters, whereas Si-based detector lengths must be long enough to ensure good X-ray absorption. In one example of Si-based detector, the Si wafers are mounted sideways or edge-on against incoming X-rays to form a deep Si strip detector~\cite{Bornefalk_2010}. Therefore, building a full-area Si detector system can be more challenging. For imaging performance, both types of \ac{PCD} have advantages and disadvantages in terms of signal quality as well as detection efficiency. More detailed comparisons can be found in~\cite{Ballabriga_2016,Danielsson_2021}. 

The innovation of \ac{PCD} makes \ac{PCCT} more attractive and offers unique advantages over conventional \ac{CT} or \ac{DECT}. These include improved dose efficiency by elimination of electronic noise, improved \ac{CNR} ratio through energy weighting~\cite{Willemink2018-oq, Samei2020-rf,Danielsson_2021}, higher spatial resolution due to the small sub-millimeter \ac{PCD} fabricated without any septa~\cite{siemens_pcct_white_paper,Danielsson_2021}, and most importantly, unprecedented material decomposition capabilities potentially for multi-tracer studies. Although \ac{PCCT} is potentially more advantageous, it has to deal with technical challenges, including charge sharing and pile-up effects together with the need for substantial hardware and system research and development. Currently, the accessibility of \ac{PCCT} for clinical applications is still limited.

\section{Multi-Energy Image Reconstruction}\label{sec:recon}

Spectral \ac{CT}, i.e., \ac{DECT} and \ac{PCCT}, offer the possibility to perform separate measurements, each measurement corresponding to an energy spectrum. One possibility is to reconstruct several attenuation \ac{CT} images at different energies from these binned raw data. These images can then be used, e.g., for image-based material decomposition \cite{niu2014iterative,feng2020image} as illustrated in the top path of Fig.~\ref{fig:tasks}; more sophisticated method, in particular the one-step reconstruction of material images, will be discussed in Section~\ref{sec:mat_decomp}.

The acquired  projections usually suffer from low \ac{SNR} due to limited photons  in each energy bin \cite{shikhaliev2009projection}. Moreover, practical constraints such as a reduced scanning time restrict \ac{CT} systems to have a limited number of views.  Therefore, the development of specific multi-energy reconstruction algorithms is of major importance.

This section reviews existing reconstruction algorithms for multi-energy \ac{CT} reconstruction from energy-binned projection data, starting from conventional \ac{CT} reconstruction algorithms to synergistic multi-energy \ac{CT} reconstruction, with the incorporation of \ac{DL} techniques and deep learning architectures. The methods presented here are only a subset of the literature in multichannel image reconstruction and we refer the readers to \citeauthor{arridge2021overview} \cite{arridge2021overview} for an exhaustive review.

\subsection{Forward and Inverse Problems}

In this section, we briefly introduce a forward model that can be equally used for \ac{PCCT} and \ac{DECT}. We consider a standard discrete model used in \ac{MBIR}.

The linear attenuation image takes the form of a spatially- and energy-dependent function $\mu \colon \R^n \times \R^+ \to \R^+$, $n=2,3$, such that for all $\boldr\in\R^n$ and for all $E\in \R^+$, $\mu(\boldr,E)$ is the linear attenuation at position $\boldr$ and energy $E$. Standard \ac{CT} systems perform measurements along a collection of rays $\{\calL_i\}$ where $\calL_i \subset \R^n$ denotes the $i$\th{} ray, $i=1,\dots,I$, with $I=N_\rmd \times N_\rms$, $N_\rmd$ and $N_\rms$ being respectively the number of detector pixels and the number of source positions. For all $i=1,\dots, I$, the expected signal (e.g. the number of photons in PCCT) is given by the Beer-Lambert law as   
\begin{equation}\label{eq:beer-lambert}
	\ybar_i(\mu) = \int_0^{+\infty} h_i(E) \cdot \rme^{ -\int_{\calL_i} \mu(\boldr,E) \,\rmd \boldr    } \, \rmd E  + r_i
\end{equation}	 
where `$\int_{\calL_i}$' denotes the line integral along $\calL_i$, $h_i$ is the corresponding X-ray photon flux which accounts for the source spectrum and the detector sensitivity (times the energy with energy integrating detectors) and $r_i$ is the background term (e.g., scatter, dark current).

In multi-energy \ac{CT} (\eg{} \ac{PCCT} and \ac{DECT}), the measurements are regrouped into $K$ energy bins ($K=2$ for \ac{DECT} and more for \ac{PCCT}). For each bin $k$, the expected number of detected X-ray photons is
\begin{equation}\label{eq:beer-lambert_binned}
	\ybar_{i,k}(\mu) = \int_0^{+\infty} h_{i,k}(E) \cdot \rme^{ -\int_{\calL_{i,k}} \mu(\boldr,E) \,\rmd \boldr   } \, \rmd E  +   r_{i,k}
\end{equation}	
where $\calL_{i,k}$ is the $i$\th{} ray for bin $k$,  $h_{i,k}$ is the photon flux X-ray intensity for bin $k$ and $r_{i,k}$ is the background term. In \ac{PCCT} each bin $k$ corresponds to an interval $[E_{k-1},E_k]$ with $E_0<E_1<\dots<E_K$, although $h_{i,k}$ may spillover the neighboring intervals. We assume that the number of detector pixels is equal to $I$ for each energy bin $k$.

The forward model \eqref{eq:beer-lambert_binned} applies to both \ac{PCCT} and \ac{DECT}. In \ac{PCCT}, the detector records the deposited energy in each interaction and the energy binning is performed the same way for each ray so that $\calL_{i,k}$ is independent of the bin $k$. In contrast, \ac{DECT} systems (except dual-layer detectors) perform 2 independent acquisitions with 2 different photon flux X-ray intensity $h_{i,1}$ and $h_{i,2}$, possibly at different source locations (i.e., via rapid \ac{kVp} switching) so that the rays generally depend on $k$.

One of the possible tasks in \ac{PCCT} and \ac{DECT} is to estimate a collection of $K$ attenuation \ac{CT} images, i.e., one image per each of the $K$ binned measurements  $\{\boldy_k\}$, $\boldy_k = [y_{1,k},\dots,y_{I,k}]\transp \in \R^I$. The energy-dependent image to reconstruct is sampled on a grid of $J$ pixels, assuming that $\mu$ can be decomposed on a basis of $J$ ``pixel-functions'' 
$u_j$ such that
\begin{equation}\label{eq:decomp}
	\mu(\boldr,E) = \sum_{j=1}^J \mu_j(E) u_j(\boldr) \, , \quad \forall (\boldr,E)\in\R^n\times\R^+ 
\end{equation} 
where $\mu_j(E)$ is the energy-dependent attenuation at pixel $j$. The line integrals in Eq.~\eqref{eq:beer-lambert} and Eq.~\eqref{eq:beer-lambert_binned} can be therefore rewritten as
\begin{equation}\label{eq:syst_mat}
	\int_{\calL_{i,k}} \mu(\boldr,E) \,\rmd \boldr  = [\boldA_k \boldmu(E)]_i 
\end{equation} 
with $\boldA_k \in \R^{I\times J}$ defined as $[\boldA_k]_{i,j} = \int_{\calL_{i,k}} u_j(\boldr) \,\rmd \boldr$ and $\boldmu(E) = [\mu_1(E),\dots,\mu_J(E)]\transp \in \R_+^J$ is the discretized energy-dependent attenuation, and we consider the following model which is an approximate version of Eq.~\eqref{eq:beer-lambert_binned} 
\begin{equation}\label{eq:beer-lambert_discrete}
	\ybar_{i,k}(\boldmu_k) =  \hbar_{i,k} \cdot \rme^{-[\boldA_k \boldmu_k]_i}
\end{equation}
where  $\hbar_{i,k} = \int_0^{+\infty}h_{i,k}(E)\rmd E$ and for each $k=1,\dots,K$  the image $\boldmu_k = [\mu_{1,k},\dots,\mu_{J,k}]\transp \in \R^J $ is an ``average'' attenuation image corresponding to energy bin $k$. 

The reconstruction of each $\boldmu_k$ is achieved by ``fitting'' the expectation $\boldybar_k(\boldmu_k) = [\ybar_{1,k}(\boldmu_k),\dots,\ybar_{I,k}(\boldmu_k)]\transp$ to the measurement  $\boldy_k$, for example by solving the inverse problem
\begin{equation}\label{eq:inv_prob}
	\boldA_k \boldmu_k = \boldb_k 
\end{equation}
with respect to $\boldmu_k$, where $\boldb_k = [b_{1,k}, \dots, b_{I,k}]\transp$, $b_{i,k} = \log \hbar_{i,k}/y_{i,k}$, is the vector of the approximated line integrals. This can be achieved by using an analytical method such as \ac{FBP} \cite{Natterer2001mathematics}, or by using an iterative technique \cite{gordon1970algebraic,gilbert1972iterative}. Unfortunately, the inverse problem~\eqref{eq:inv_prob} is ill-posed and direct inversion leads to noise amplification which is impractical for low-dose imaging. Moreover, the inversion relies on an idealized mathematical model that does not reflect the physics of the acquisition, especially by ignoring the polychromatic nature of the X-ray spectra.

\subsection{Penalized Reconstruction}\label{sec:pen_recon}

Alternatively, the reconstruction can be achieved for each energy bin $k$ by finding an estimate $\boldmuhat_k$ as the solution of an optimization problem of the form 
\begin{equation}\label{eq:recon}
	\boldmuhat_k	\in	 \argmin_{\boldmu_k\in \R^J_+} \, L\left(\boldy_k , \boldybar_k(\boldmu_k)\right) + \beta_k R_k(\boldmu_k)  
\end{equation} 
where $L$ is a loss function (e.g., the Poisson negative log-likelihood for \ac{PCCT}) that evaluates the goodness of fit between the data $\boldy_k$ and $\boldybar_k(\boldmu_k)$, $\beta_k>0$ is a weight and $R_k$ is a penalty function or regularizer, generally convex and nonnegative, that promotes desired image properties while controlling the noise. The data fidelity term  in \eqref{eq:recon} is convex when $r_{i,k} = 0$ for all $i,k$. Although many approaches were proposed to solve \eqref{eq:recon}, most algorithms are somehow similar to the proximal gradient algorithm \cite{beck2009fast,combettes2011proximal}, that is to say, given an image estimate $\boldmu_k^{(q)}$ at iteration $q$, the next estimate $\boldmu_k^{(q+1)}$ is obtained via a reconstruction step followed by a smoothing step,
\begin{align}
	\boldmu_k^{(q + 1/2)} &  = {} \boldmu_k^{(q)} - \boldH_k^{-1} \boldg_k^{(q)} \label{eq:gradient} \\
	\boldmu_k^{(q + 1)}   &  = {} \argmin_{\boldmu_k \in \R^J_+} \, \frac{1}{2} \left\| \boldmu_k - \boldmu_k^{(q + 1/2)} \right\|^2_{\boldH_k} + \beta_k R_k(\boldmu_k) \, ,\label{eq:smoothing}
\end{align}	
where $\boldg_k^{(q)}$ is the gradient of the data fidelity loss $\boldmu_k\mapsto L\left(\boldy_k , \boldybar_k(\boldmu_k)\right)$ evaluated at $\boldmu_k^{(q)}$ and $\boldH_k$ is a suitable diagonal positive-definite matrix (typically, a diagonal majorizer of the Hessian of the data fidelity loss).    The first step \eqref{eq:gradient} is a gradient descent that guarantees a decrease of the data fidelity  while the second step \eqref{eq:smoothing} is an image denoising operation. This type of approach encompasses optimization transfer techniques such as \ac{SQS} \cite{erdogan1999monotonic,elbakri2002statistical}. 

The choice of $R_k$ depends on the desired image properties. A popular choice consists in  penalizing differences in the values of neighboring pixels with a smooth edge-preserving potential function and solving Eq.~\eqref{eq:smoothing} is achieved with standard smooth optimization tools \cite{erdogan1999monotonic,elbakri2002statistical}. Another popular choice  is the \ac{CS} approach, which has been widely used in medical imaging when using an undersampled measurement operator $\boldA_k$ (e.g., sparse-view \ac{CT}).  \Ac{CS} consists of assuming that the signal to recover is sparse in some sense  to recover it from far fewer samples than required by the Nyquist–Shannon sampling theorem. In the following paragraphs, we briefly discuss the synthesis and the analysis approaches.

In the synthesis approach, it is assumed that $\boldmu_k = \boldD_k \boldz_k$ where $\boldD_k \in \R^{J\times S}$ is a dictionary matrix, i.e., an over-complete basis, consisting of $S$ atoms, and $\boldz_k \in \R^S$ is a sparse vector of coefficients such that $\boldmu_k$ is represented by a fraction of columns of $\boldD_k$, or atoms. The reconstruction of the image is then given by
\begin{align}
    \boldzhat_k & = {} \argmin_{\boldz_k \in \R^S} \, L\left( \boldy_k, \boldybar_k(\boldD_k \boldz_k) \right) + \alpha \|\boldz_k\|_m \nonumber \\
	\boldmuhat_k & = {} \boldD_k \boldzhat_k \label{eq:synthesis_model}
\end{align}   
where  $\|\cdot\|_m$ can be either the $\ell^0$ semi-norm or its convex relaxation, the $\ell^1$ norm, and $\alpha>0$ is a weight controlling the sparsity of $\boldz$. The optimization can be achieved by orthogonal matching pursuit \cite{pati1993orthogonal} for $m=0$ and proximal gradient for $m=1$. In some situations, imposing  $\boldmu_k = \boldD_k \boldz_k$ may be too restrictive and a relaxed constraint $\boldmu_k \approx \boldD_k \boldz_k$ is often preferred. The reconstruction is then achieved by penalized reconstruction using a regulariser $R_{\boldD_k}$ that prevents $\boldmu_k$ from deviating from $\boldD_k \boldz_k$, usually defined as
\begin{equation}\label{eq:penalty_synthesis}
	R_{\boldD_k}(\boldmu_k) = \min_{\boldz_k \in  \R^S} \, \frac{1}{2} \|\boldmu_k - \boldD_k \boldz_k \|^2_2 + \alpha_k \|\boldz_k\|_m \, .
\end{equation}
where $\alpha_k>0$ is a weight. Solving Eq.~\eqref{eq:recon} is achieved by alternating between minimization in $\boldmu_k$ (e.g., by performing several iterations of  \eqref{eq:gradient} and \eqref{eq:smoothing}) and  minimization in $\boldz_k$ (e.g., orthogonal matching pursuit \cite{pati1993orthogonal} for $m=0$ and proximal gradient for $m=1$). This type of penalty forms the basis of learned penalties that we will address in Section~\ref{sec:data_learned_syn_pen}.

In the analysis (encoding) approach, it is assumed that $\boldT_k \boldmu_k$ is sparse, where $\boldT_k \in \R^{D\times J}$ is a sparsifying transform, and the penalty $R_k$ is 
\begin{equation}\label{eq:cs}
	R_{\boldD_k}(\boldmu_k) = \|\boldT_k \boldmu_k \|_m \, 
\end{equation}
For example, in image processing, $\boldT_k$ can be a wavelet transform  or finite differences (discrete gradient). In the latter case and when $m=1$, the corresponding penalty $R_k$ is referred to as \ac{TV}\footnote{This definition of \ac{TV} corresponds to \emph{anisotropic} \ac{TV}. The alternative form, \emph{isotropic} \ac{TV} consists of summing the $\ell^2$-norm of the gradient at each pixel, and had been widely used in \ac{CT} reconstruction \cite{ritschl2011improved,sidky2012convex,liu2013total,niu2014sparse}. Both \ac{TV} penalties can be addressed  by proximal gradient \cite{beck2009fast2}. Alternatively, the $\ell^0$ semi-norm can also be used \cite{xu2011image}.}. \Ac{TV} has been extensively used in image processing for its ability to represent piecewise constant objects \cite{strong2003edge}. Because $R_{\boldD_k}$ is non-smooth, solving Eq.~\eqref{eq:smoothing} requires variable splitting techniques such as proximal gradient,  \ac{ADMM} \cite{Boyd2010} or the Chambolle-Pock algorithm \cite{chambolle2011first}.

\subsection{Synergistic Penalties}\label{sec:syn_recon}

Alternatively, the images can be simultaneously reconstructed. Introducing $\boldmu = \{ \boldmu_k\}$ the spectral \ac{CT} multichannel image, $\boldy = \{\boldy_k\}$ the binned projection data and $\boldybar(\boldmu) = \{ \boldybar_k(\boldmu_k)\}$ the expected binned projections, the images can be simultaneously reconstructed as
\begin{equation}\label{eq:synergistic_recon}
	\boldmuhat \in  \argmin_{\boldmu} \,  L\left(\boldy , \boldybar(\boldmu) \right) + \beta R(\boldmu)    
\end{equation}
where $R$ is a \emph{synergistic} penalty function that promotes structural and/or functional dependencies between the multiple images and a proximal gradient algorithm to solve Eq.~\eqref{eq:synergistic_recon} at iteration $q+1$ to update $\boldmu^{(q)} = \{\boldmu_k^{(q)}, \, k=1,\dots,K\}$ is 
\begin{align}
	\boldmu_k^{(q + 1/2)} &  = {}\boldmu_k^{(q)} - \boldH_k^{-1} \boldg_k^{(q)} \label{eq:gradient_k},\quad \forall k \\
	\boldmu^{(q + 1)} &  = {} \argmin_{\boldmu } \, \sum_{k=1}^K \frac{1}{2} \left\| \boldmu_k - \boldmu_k^{(q + 1/2)} \right\|^2_{\boldH_k} + \beta R(\boldmu) \, ,\label{eq:smoothing_syn}
\end{align}	
where Eq.~\eqref{eq:smoothing_syn} corresponds to a synergistic smoothing step. The paradigm shift here is that allowing the channels to ``talk to each other'' can reduce the noise as each channel participates in the reconstruction of all the other ones. In the context of spectral \ac{CT}, this suggests that the reconstruction of each image $\boldmu_k$ benefits from the entire measurement data $\boldy$. Here, we present a non-exhaustive list of existing approaches. 

One class of approaches consists of enforcing structural similarities between the $K$ channels. Examples include \ac{JTV} which encourages gradient-sparse solutions (in the same way as the conventional \ac{TV}) and also encourages joint sparsity of the gradients  \cite{sapiro1996anisotropic,blomgren1998color}. \Ac{TNV} encourages common edge locations and a shared gradient direction among image channels \cite{lefkimmiatis2013convex,rigie2015joint}. All these works reported improved image quality with synergistic image processing as compared with single-image processing.

A second class of approaches consists of promoting similarities across channels by controlling the rank of the multichannel image. Given that the energy dependence of human tissues can be represented by the linear combination of two materials only (see Section~\ref{sec:mat_decomp}), it is natural to expect a low rank in some sense in the spectral dimension. For dynamic \ac{CT} imaging, \citeauthor{gao2011robust}~\cite{gao2011robust} proposed a method, namely Robust Principle Component Analysis based 4-D CT (RPCA-4DCT), based on a \ac{LR} + sparse decomposition of the multichannel image matrix $\boldM = [\boldmu_1,\dots,\boldmu_K]\in \R^{J\times K}$ ($K$ time frames), 
\begin{equation}
	\boldM = \boldM_\mathrm{l} + \boldM_\mathrm{s}
\end{equation}
where $\boldM_\mathrm{l}$ is an \ac{LR} matrix representing the information that is repeated across the channels and $\boldM_\mathrm{s}$ is a sparse matrix representing the variations in the form of outliers, and a synergistic penalty defined as  
\begin{equation}
	R(\boldM) = \gamma\|\boldM_\mathrm{l}\|_* + \|\boldM_\mathrm{s}\|_1
\end{equation}
$\gamma>0$ and the nuclear norm $\|\cdot\|_*$ is a relaxation of the rank of a matrix, and showed that their approach outperforms \ac{TV}-based (in both spatial and temporal dimensions) regularization.  \citeauthor{gao2011multi}~\cite{gao2011multi} then generalized this method for spectral \ac{CT} with the \ac{PRISM}, which uses the rank of a tight-frame transform of the \ac{LR} matrix to better characterize the multi-level and multi-filtered image coherence across the energy spectrum, in combination  with energy-dependent intensity information, and showed their method outperformed  conventional \ac{LR} + sparse decomposition. This principle was further generalized by ``folding'' the multichannel image $\boldM\in \R^{J\times K}$ in a 3-way tensor $\calM \in \R^{\sqrt{J}\times \sqrt{J}\times K}$ (for \ac{2D} imaging) and applying the generalized tensor nuclear norm regularizer to exploit structural redundancies across spatial dimensions (in addition to the spectral dimension) \cite{chu2012multi,semerci2014tensor,li2014tensor,li2014spectral,xia2019spectral,he2023spectral}.

A third and different class of approaches consists of enforcing structural similarities of each $\boldmu_k$ with a reference low-noise high-resolution image $\bar{\boldmu}$, generally taken as the reconstruction from all combined energy bins. Instead of using a joint penalty $R$, each channel is controlled by a penalty $R_k$ of the form
\begin{equation}\label{eq:piccs}
	R_k(\boldmu_k) = S_k(\boldmu_k,\bar{\boldmu})
\end{equation} 
where $S$ is a ``similarity measure'' between $\boldmu_k$ and the reference image $\bar{\boldmu}$. The \ac{PICCS} \cite{chen2008prior,yu2016spectral} approach uses $S(\boldmu_k,\bar{\boldmu}) = \| \nabla (\boldmu_k - \bar{\boldmu})\|_m$, $\nabla$ denoting the discrete gradient; the $\ell^1$-norm can also be replaced with the $\ell^0$ semi-norm \cite{wang2020low}. Variants of this approach include nonlocal similarity measures \cite{zhang2016spectral,yao2019multi} to preserve  both high- and low-frequency components. More recently, \citeauthor{cueva2021synergistic} \cite{cueva2021synergistic} proposed the directional \ac{TV} approach for spectral \ac{CT}, which enforces colinearity  between the gradients of $\boldmu_k$  and $\bar{\boldmu}$, while preserving sparsity, and showed their approach outperforms \ac{TV}.   

To conclude, spectral \ac{CT} reconstruction with synergistic penalties has been widely used  to improve the quality of the reconstructed images. However, the success of this approach heavily depends on the selection of an appropriate synergistic penalty term, which is typically fixed and may not always accurately reflect the true underlying structure of the data.

\subsection{Learned Penalties}\label{sec:data_learned_syn_pen}

Traditional regularization methods, such as those described in Sections~\ref{sec:pen_recon} and~\ref{sec:syn_recon}, impose a fixed handcrafted penalty on the reconstructed image based on certain assumptions about its structure, such as sparsity or smoothness. However, these assumptions may not always hold in practice, leading to suboptimal reconstructions. Learned penalty functions, on the other hand, can adaptively adjust the penalty term based on the specific characteristics of the data, allowing for more accurate and flexible reconstruction. 

This subsection discusses learned synergistic penalties for multichannel image reconstruction. In particular, we will focus on penalties based on a \emph{generator} $\boldcalG$, which is a trained mapping that takes as input a latent variable $\boldz$, which can be an image or a code, and returns a plausible multichannel image  $\boldcalG(\boldz) = \{ \boldcalF_k(\boldz) \}$. The latent variable $\boldz$ represents the patient which connects the different channels. The penalty function plays the role of a \emph{discriminator} by promoting images originating from the generative model and by penalizing images that deviate from it, in a similar fashion to the relaxed synthesis model \eqref{eq:penalty_synthesis}.

Most of this subsection will address \ac{DL}, i.e., $\boldcalF_k(\boldz) = \boldD_k \boldz$ for some dictionary matrix $\boldD_k$, as it is the most prevalent learned penalty used in synergistic multichannel image reconstruction. \Ac{CDL} will also be discussed in a short paragraph. Finally, we will discuss recent work that uses deep \ac{NN} models.

In this subsection $\boldmu^\mathrm{tr} = \left\{\boldmu_k^\mathrm{tr}\right\} $ denotes a random spectral \ac{CT} image whose joint distribution corresponds to the empirical distribution derived from a training dataset  of $L$ spectral \ac{CT} images $\boldmu^{\mathrm{tr},[1]},\dots,\boldmu^{\mathrm{tr},[L]}\in \left(\R^J\right)^K$, that is to say for all mapping $h \colon \left(\R^J\right)^K \to \R$,
\begin{equation}
	\bbE\left[ h \left(  \boldmu^\mathrm{tr}   \right)\right]  = \frac{1}{L} \sum_{\ell=1}^L h \left(   \boldmu^{\mathrm{tr},[\ell]} \right) \, .
\end{equation}

\subsubsection{Dictionary Learning}\label{sec:dict}

For simplicity this section will consider \ac{2D} imaging (i.e., $n=2$), so that each image $\boldmu_k\in\R^J$ can be reshaped into a $\sqrt{J} \times \sqrt{J}$ square matrix.

\Ac{DL} is a popular technique for regularizing the reconstruction process in medical imaging and especially in \ac{CT} reconstruction \cite{xu2012low,zhang2016low,komolafe2020smoothed,xu2020limited}. The basic idea behind \ac{DL} is to learn a dictionary matrix that can represent the image with a fraction of its columns. The dictionary operator requires a large number of atoms to accurately represent all possible images which increase the  computational complexity of training. Therefore,  to reduce the complexity, the image is generally split into $P$ smaller $d$-dimensional ``patches'' (possibly overlapping) with $d\ll J$. For a given energy bin $k$, the trained penalty to reconstruct a single attenuation image $\boldmu_k$ by penalized reconstruction \eqref{eq:recon} is given by
\begin{equation}\label{eq:penalty_synthesis_patches}
	R_{\boldD^\star_k}(\boldmu_k) = \min_{\{\boldz_p\}} \, \sum_{p=1}^P\frac{1}{2} \|\boldP_p\boldmu_k - \boldD_k^\star \boldz_p \|^2_2 + \alpha \|\boldz_p \|_m
\end{equation}
where $\boldD^\star_k \in \R^{d\times S}$ is the trained dictionary matrix, $\boldP_p \in \R^{d\times J}$ is the $p$\th{} patch extractor and each $\boldz_p$ is the sparse vector of coefficients to represent the $p$\th{} patch with $\boldD^\star_k $. The training is generally performed by minimizing $R_{\boldD_k}$ with respect to $\boldD_k$ (with unit $\ell^2$-norm constraints on its columns) over a training data set of high-quality images,
\begin{equation}
	\boldD^\star_k = \argmin_{\boldD_k} \, \bbE \left[ R_{\boldD_k}\left(\boldmu^{\mathrm{tr}}_{k}\right)  \right]
\end{equation}
for example using the K-SVD algorithm introduced by \citeauthor{aharon2006k} \cite{aharon2006k}.

\Ac{DL} can also be used to represent images synergistically. \Ac{TDL} consists in folding the spectral images $\boldmu = \{\boldmu_k\}\in \left(\R^J\right)^K$  into a tensor $\calM \in \R^{\sqrt{J} \times \sqrt{J} \times K}$ and in training a  spatio-spectral tensor dictionary to sparsely represent $\calM$ with a sparse \emph{core tensor} $\calZ \in \R^{s_1 \times s_2 \times s_3}$, such that each atom conveys information across the spectral dimension. A common approach used to sparsely represent the sensor image $\calM$ is to use the Tucker decomposition \cite{tucker1966some,zubair2013tensor}. It was utilized in multispectral image denoising \cite{liu2012denoising,peng2014decomposable} as well as in dynamic \ac{CT} \cite{tan2015tensor} (by replacing the spectral dimension by the temporal dimension). Denoting $\calP_p \colon \R^{\sqrt{J} \times \sqrt{J} \times K} \to \R^{\sqrt{d} \times \sqrt{d} \times K}$ the $p$\th{} spatio-spectral image patch extractor, each patch $\calP_p(\calM)$ can be approximated by the Tucker decomposition as
\begin{equation}\label{eq:tucker_decomp}
	\calP_p(\calM) \approx \calZ_p \times_1 \boldD^{(1)} \times_2 \boldD^{(2)} \times_3 \boldD^{(3)}
\end{equation}
where $\calZ_p \in \R^{s_1 \times s_2 \times s_3}$ is the core tensor for the $p$\th{} patch, $\boldD^{(1)} \in \R^{\sqrt{d}\times s_1}$ and $\boldD^{(2)} \in \R^{\sqrt{d}\times s_2}$ are the \ac{2D} spatial dictionaries along each dimension and $\boldD^{(3)} \in \R^{K\times s_3}$ is the spectral dictionary (all of them consisting of orthogonal unit column vectors), and $\times_n$ is the mode-$n$ tensor/matrix product (see for example  \citeauthor{semerci2014tensor}~\cite{semerci2014tensor} for a definition of tensor-matrix product).

The Tucker decomposition requires a large number of atoms and therefore is cumbersome for \ac{DL} in high dimensions. To remedy this, \citeauthor{zhang2016tensor}~\cite{zhang2016tensor} proposed to use the  \ac{CPD}, which consists of assuming that the core tensor $\calZ$ is diagonal, i.e., $s_1=s_2=s_3=S$ and $(\calZ)_{a,b,c}\ne 0 \implies a=b=c$, which leads to the following approximation \cite{zubair2013tensor},
\begin{equation}\label{eq:cpd}
	\calP_p(\calM) \approx  \sum_{s=1}^S z_{s,p} \, \calD_s \, ,
\end{equation}
where for all $s$, $\calD_s = \boldd^{(1)}_s \otimes \boldd^{(2)}_s \otimes \boldd^{(3)}_s \in \R^{\sqrt{d} \times \sqrt{d} \times K}$, $\boldd^{(1)}_s,\boldd^{(2)}_s\in\R^{\sqrt{d}}$ and $\boldd^{(3)}_s \in \R^K$ are unit vectors, $\boldz_p = [z_{1,p},\dots,z_{S,p}]\transp\in\R^S$ is a sparse vector corresponding to the diagonal of $\calZ_p$ and `$\otimes$' denotes the matrix outer product. \citeauthor{zhang2016tensor} then used this decomposition to train spatio-spectral dictionaries combined with a K-\ac{CPD} algorithm \cite{duan2012k} from which the following penalty term is derived\footnote{In \cite{zhang2016tensor}, \citeauthor{zhang2016tensor}~trained zero-mean atoms and therefore subtracted a channel-mean from each patch in Eq.~\eqref{eq:cpd_penalty}. }:
\begin{align}
	R_{\calD^\star}(\calM)  = & \min_{\{\boldz_p\}} \, \sum_{p=1}^P \frac{1}{2} \left\|  \calP_p(\calM) -  \sum_{s=1}^S z_{s,p} \, \calD^\star_s    \right\|^2_\mathrm{F} + \alpha \| \boldz_{p} \|_m \, \label{eq:cpd_penalty}
\end{align}
with $\calD^\star = \{\calD^\star_s\}$. The training is performed as 
\begin{equation} 
	\calD^\star = \argmin_{\calD} \, \bbE \left[  R_{\calD}(\calM^\mathrm{tr}) \right] 
\end{equation}
where $\calM^\mathrm{tr}$ is the spatio-spectral tensor obtained by folding the $n$\th{} training multichannel image matrix $\left[\boldmu_1^\mathrm{tr},\dots,\boldmu_{K}^\mathrm{tr}\right]$, and the minimization is performed subject to the constraint $\calD_s = \boldd^{(1)}_s \otimes \boldd^{(2)}_s \otimes \boldd^{(3)}_s$. \citeauthor{wu2018low}~\cite{wu2018low} proposed a similar approach with the addition of the $\ell^0$ semi-norm of the gradient images at each energy bin in order to enforce piecewise smoothness of the images, while \citeauthor{li2022tensor}~\cite{li2022tensor} added  a \ac{PICCS}-like penalty \eqref{eq:piccs} to enforce joint sparsity of the gradients. 

We can observe that the \ac{TDL} regularizer with \ac{CPD} can be rewritten as 
\begin{equation}\label{eq:penalty_synthesis_patches_mult}
	R_{\calD^\star}(\boldmu) =  \min_{\{\boldz_p\}} \,\sum_{p=1}^P \sum_{k=1}^K   \frac{1}{2}   \left\| \boldP_p \boldmu_k  - \boldD^\star_k \boldz_p   \right\|_2^2  +  \alpha \|\boldz_p\|_m
\end{equation}
where each column of $\boldD^\star_k \in \R^{d\times S}$ is the matrix $ \left[\boldd^{(3)\star}_s\right]_k \cdot \left(\boldd^{(1)\star}_s \otimes \boldd^{(2)\star}_s\right)$ reshaped into a vector. This regularizer is a generalization of \eqref{eq:penalty_synthesis_patches} to multichannel imaging with a collection of dictionaries $\{\boldD_k^\star\}$ and a unique sparse code $\{\boldz_p\}$ for all energy bins $k$. Similar representations were used in coupled \ac{DL} in multimodal imaging synergistic reconstruction, such as in \ac{PET}/\ac{MRI} \cite{sudarshan2018joint,sudarshan2020joint}, multi-contrast \ac{MRI} \cite{song2019coupled} as well as super-resolution \cite{yang2012coupled}.

Patch-based \ac{DL} may be inefficient as the atoms are shift-variant and may produce atoms that are shifted versions of each other. Moreover, using many neighboring/overlapping patches across the training images is not efficient in terms of sparse representation as sparsification is performed on each patch separately.  Instead, \ac{CDL} \cite{wohlberg2015efficient,chun2017convolutional,garcia2018convolutional} consists in utilizing a trained dictionary of image filters to represent the image as a linear combination of sparse feature images convolved with the filters (synthesis model) that can be used in a penalty function similar to Eq.~\eqref{eq:penalty_synthesis_patches}, without patch extraction. \citeauthor{bao2019convolutional}~\cite{bao2019convolutional} used this approach for \ac{CT} \ac{MBIR}. Alternatively, \ac{CAOL} consists in training sparsifying convolutions, which can then be used as a penalty function for \ac{MBIR} \cite{chun2019Convolutional2}. There are a few applications of \ac{CDL} and \ac{CAOL} in multichannel imaging and multi-energy \ac{CT} (see \cite{garcia2018convolutional2} for a review). \citeauthor{degraux2017online}~\cite{degraux2017online} proposed a multichannel \ac{CDL} model to represent two images simultaneously (intensity-depth imaging), using a collection of pairs of image filters.  \citeauthor{gao2022multi}~\cite{gao2022multi} proposed a more general model with common  and unique filters. More recently, \citeauthor{perelli2022multi} \cite{perelli2022multi} proposed a multichannel \ac{CAOL} for \ac{DECT} joint reconstruction, which uses pairs of image filters to jointly sparsify the low- and high-energy images, and demonstrated their method outperforms \ac{JTV}-based synergistic reconstruction. 

\subsubsection{Deep-Learned Penalties}

The synthesis model used in \ac{DL} can be generalized by replacing the multichannel dictionaries $\{\boldD_k\}$ with a trained multi-branch \ac{NN} $\boldcalG_{\boldtheta^\star} (\boldz) = [ \boldcalF_{\boldtheta_1^\star}(\boldz) , \dots, \boldcalF_{\boldtheta_K^\star}(\boldz)]$  which maps a single input $\boldz$ to a collection of images $\{\boldcalF_{\boldtheta_k^\star}(\boldz)\}$, $\boldtheta^\star = \{\boldtheta_k^\star\}$ designed to represent the spectral \ac{CT} image $\boldmu = \{\boldmu_k\}$. Unlike dictionary learning, which uses a finite number of atoms to represent the data, deep \acp{NN} can learn parameters that can capture more intricate patterns and structures in the image data. A synergistic regularizer used in Eq.~\eqref{eq:recon} can then be defined as 
\begin{equation}\label{eq:gen_model_syn}
	R_{\boldtheta^\star}(\boldmu) = \min_{\boldz} \, \sum_{k=1}^K\left\|  \boldmu_k - \boldcalF_{\boldtheta_k^\star}(\boldz)  \right\|_2^2 + \alpha H(\boldz)
\end{equation}
where $H$ is a penalty function for $\boldz$ (not necessarily sparsity-promoting), which is the generalization of multichannel \ac{DL} \eqref{eq:penalty_synthesis_patches_mult} using multiple \acp{NN}. \citeauthor{wang2022synergistic}~\cite{wang2022synergistic} used this approach with a collection of U-nets $\boldcalF_{\boldtheta_k}$ trained in a supervised way to map the attenuation image at the lowest energy bin $\boldmu_1$ to the attenuation image at energy bin $k$, i.e., 
\begin{equation}
	\boldtheta_k^\star  = \argmin_{\boldtheta_k}  \,  \bbE \left[ \left\| \boldmu_{k}^\mathrm{tr} - \boldcalF_{\boldtheta_k} \left(\boldmu_{1}^\mathrm{tr}\right)  \right\|^2_2 \right]\quad \forall k  
\end{equation}	
and combined a standard Huber penalty   (the $H$ function in Eq.~\eqref{eq:gen_model_syn}) for $\boldz$. The trained penalty $R_{\boldtheta_k^\star}$ ``connects'' the channels by a spectral image   $\{\boldmu_k\}$ such that each $\boldmu_k$ originates from a single image $\boldz$ that is smooth in the sense of $H$. \citeauthor{wang2022synergistic} reported substantial noise reduction as compared with individually reconstructed images and \ac{JTV} synergistic reconstruction.

The training of the generative model can also be unsupervised, for example as a multichannel \ac{AE}, i.e.,
\begin{equation}\label{eq:ae}	
	\boldtheta^\star =   \argmin_{\boldtheta} \min_{\boldphi} \, \bbE \left[ \big\| \boldmu^\mathrm{tr}   
	- \boldcalG_{\boldtheta}  \left(  \boldE_{\boldphi} \left( \boldmu^\mathrm{tr}_k  ,\dots,\boldmu_{K}^\mathrm{tr}    \right)     \right)     \big\|^2_2   \right]   \nonumber 
\end{equation}
where $\boldE_{\boldphi}  \colon \left(\R^J\right)^K \to Z$, $Z$ being the latent space, is a multichannel encoder, i.e., that encodes a collection of images into a single latent vector, parametrized with $\boldphi$, and $\boldcalG_{\boldtheta} = \{\boldcalF_{\boldtheta_k}\} : Z \to \left(\R^J\right)^K $ is the multichannel decoder.  In this approach, $\boldmu_k$ is encouraged not to deviate from the ``manifold'' of plausible images  $\{ \boldcalF_{\boldtheta_k^\star}(\boldz) , \, \boldz \in Z\}$.
\citeauthor{pinton2023synergistic}~\cite{pinton2023synergistic}  and \citeauthor{gautier2023vae}~\cite{gautier2023vae} used this approach  respectively for \ac{PET}/\ac{CT} and \ac{PET}/\ac{MRI} using a multi-branch variational \ac{AE}, and reported considerable noise reduction by reconstructing the images synergistically as opposed to reconstructing the images individually. A patched-based version of this penalty with a K-Sparse~\ac{AE} (i.e., with $H = \|\cdot\|_0$) was proposed by \citeauthor{wu2017iterative}~\cite{wu2017iterative} for  single-channel \ac{CT}. \citeauthor{duff2021regularising}~\cite{duff2021regularising} proposed a similar approach with a \ac{WGAN}.

An alternative approach, namely the \ac{DIP} introduced by \citeauthor{ulyanov2018deep}~\cite{ulyanov2018deep}, consist of fixing the input $\boldz$  and to optimize with respect to $\boldtheta$, in such a way that the reconstruction does not require pre-training of the \ac{NN}. A multichannel version of this approach using a multi-branch \ac{NN} with a single input $\boldz$ was proposed for \ac{DECT} \cite{gong2019low}.

Although deep-learned penalties have been successfully applied in image reconstruction, their application to spectral \ac{CT} has been relatively limited and remains an active area of research. Future work should focus on developing more efficient and accurate deep-learned penalties that are specifically tailored to the unique challenges and opportunities of spectral \ac{CT}.

\subsection{Deep Learning-based Reconstruction}\label{sec:data_learned_syn_recon}

Another paradigm shift has been the development of end-to-end learning architectures that directly map the raw projection data to the reconstructed images. This approach, known as learned reconstruction, has two main categories: direct reconstruction and unrolling techniques. Direct reconstruction involves training a single \ac{NN} to perform the reconstruction task, while unrolling techniques aim to mimic the iterative algorithm by ``unrolling'' its iterations into layers. These techniques have shown great potential in image reconstruction, where the acquisition of data at different energy levels provides additional information about the material composition of the imaged object. In this section, we review recent advances of unrolling-based architectures for image reconstruction and their extension to synergistic spectral \ac{CT} reconstruction. Direct methods have not yet been deployed for spectral~\ac{CT} and will be discussed in  Section~\ref{sec:perspectives}.

In the following $\left( \boldmu^\mathrm{tr},\boldy^\mathrm{tr}\right) \in \left(\R^J\right)^K \times \left(\R^I\right)^K$  denotes a random  spectral \ac{CT} image/binned projections pair whose joint distribution corresponds to the empirical distribution derived from $L$ training pairs $\left( \boldmu^{\mathrm{tr},[1]},\boldy^{\mathrm{tr},[1]}\right),\dots,\left( \boldmu^{\mathrm{tr},[L]},\boldy^{\mathrm{tr},[L]}\right) \in \left(\R^J\right)^L \times \left(\R^I\right)^L$ such that for all $\ell=1,\dots,L$ the spectral \ac{CT} multichannel image $\boldmu^{\mathrm{tr},[\ell]}$ is reconstructed from $\boldy^{\mathrm{tr},[\ell]}$.

Unrolling techniques, or \emph{learned iterative schemes}, have become increasingly popular for image reconstruction in recent years, due to their ability to leverage the flexibility and scalability of deep neural networks while retaining the interpretability and adaptability of classical iterative methods. Unrolling-based techniques aim at finding a deep architecture that approximates an iterative algorithm. 

For all energy bins $k$, the $(q+1)$\th{} iteration of an algorithm to reconstruct the image $\boldmu_k$ can be written as
\begin{equation}\label{eq:unrolling}
	\boldmu_k^{(q+1)} = \boldcalL^k_{\boldtheta_{q,k}} \left(\boldmu_k^{(q)}\right)
\end{equation}
where $\boldcalL^k_{\boldtheta_{q,k}}$  is an image-to-image mapping that intrinsically depends on $\boldy_k$ and that updates the image at layer $q$ to layer $q+1$. The  parameter $\boldtheta_{q,k}$ typically comprises algorithm hyperparameters such as step lengths and penalty weights but also \ac{NN} weights. For example, Eq.~\eqref{eq:gradient} and Eq.~\eqref{eq:smoothing} are unrolled with $\boldcalL^k_{\boldtheta_{q,k}} \left(\boldmu_k^{(q)}\right) =  \mathbf{prox}^{\boldH_k}_{\beta_{q,k} R_k}\left( \boldmu_k^{(q)} - \boldH_k^{-1} \boldg_k \right) $ where $\mathbf{prox}^{\boldH}_f(\boldx) = \argmin \frac{1}{2}  \|\cdot-\boldx\|_{\boldH}^2 + f$ and $\boldtheta_{q,k} = \beta_{q,k}$. The  $Q$-layer reconstruction architecture $\boldcalR^k_{\boldtheta_k}$, $\boldtheta_k = \{\boldtheta_{q,k}\}_{q=1}^Q$, to reconstruct $\boldmu_k$ from $\boldy_k$ is given as 
\begin{equation}\label{eq:unrolling_all}
	\boldcalR^k_{\boldtheta^\star_k}(\boldy_k) = \boldcalL^k_{\boldtheta^\star_{Q,k}} \circ \dots \circ \boldcalL^k_{\boldtheta^\star_{1,k}}        \left(\boldmu_k^{(0)}\right) \,
\end{equation}
where $\boldmu_k^{(0)}$ is a given initial image and the right-hand side depends on $\boldy_k$ by means of of $\boldcalL^k_{\boldtheta_{q,k}}$, and the trained parameter $\boldtheta^\star_k$ is obtained by supervised training as
\begin{equation}\label{eq:unrolling_training}
	\boldtheta^\star_k = \argmin_{\boldtheta_k }\,  \bbE \left[  L\left( \boldcalR^k_{\boldtheta_k}\left(\boldy_k^\mathrm{tr}\right) ,  \boldmu_k^\mathrm{tr}  \right) \right]\quad \forall k \, .
\end{equation}	
Alternative to Eq.~\eqref{eq:unrolling} and \eqref{eq:unrolling_all}, for example incorporating memory from previous iterates  at each layer, can be found in \citeauthor{arridge2019solving}~\cite{arridge2019solving}. By utilizing components of iterative algorithms such as the backprojector  $\boldA_k\transp$, unrolling-based architectures can map projection data to images without suffering from scaling issues. Many works from the literature derived unrolling architecture from existing model-based  algorithms  and we will only cite a non-exhaustive list; we refer the reader to \citeauthor{monga2021algorithm}~\cite{monga2021algorithm} for a review of unrolling techniques until 2021. One of the first unrolling architectures, namely ADMM-net, was proposed by \citeauthor{yang2016deep}~\cite{yang2016deep} for \ac{CS} \ac{MRI} and consists in a modified  \ac{ADMM} algorithm \cite{Boyd2010}  where basics operation (finite-difference operator, soft-thresholding, etc.) are replaced by transformations such as convolution layers with parameters that are trained end-to-end. Other works rapidly followed for regularized inverse problems in general and  image reconstruction in particular.  Learned proximal operators, which consist of replacing the update \eqref{eq:smoothing} with a trainable \ac{CNN} \cite{meinhardt2017learning,gupta2018cnn}. In a similar fashion, \citeauthor{chun2018deep}, proposed BCD-Net \cite{chun2018deep} and its accelerated version Momentum-Net~\cite{chun2020momentum}  which consists in unrolling a variable-splitting algorithm and replace the image regularization step with a \ac{CNN}.  \citeauthor{adler2018learned}~\cite{adler2018learned} proposed a trainable unrolled version of the primal-dual (Chambolle-Pock) algorithm~\cite{chambolle2011first}. 

\begin{figure*}
	\centering
	\input{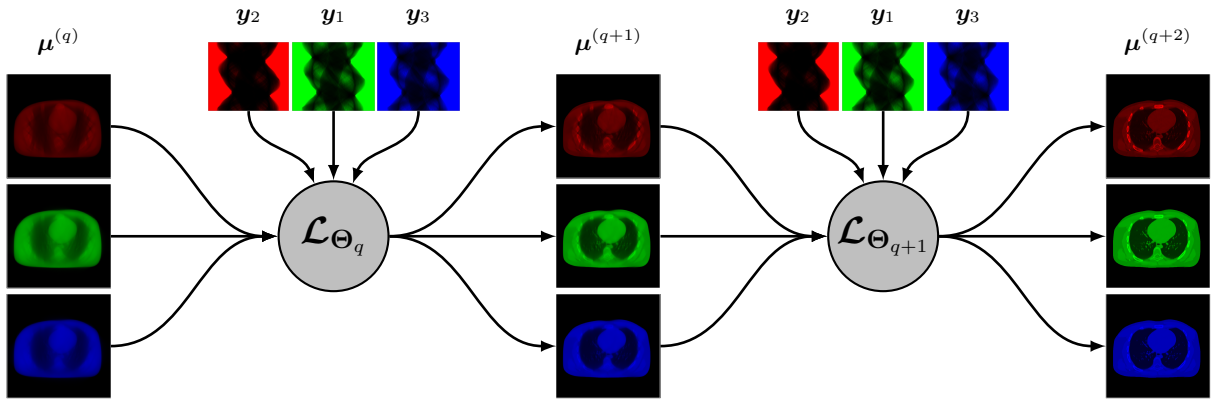}
	\caption{Representation of the synergistic unrolling architecture described in Eq.~\eqref{eq:unrolling_syn} with 3 energy bins $k=1,2,3$.  }\label{fig:unrolling}
\end{figure*}

A synergistic reconstruction algorithm such as given by Eq.~\eqref{eq:gradient_k} and Eq.~\eqref{eq:smoothing_syn} may also be unrolled in a trainable deep multi-branch architecture by merging the mappings $\boldcalL^k_{\boldtheta_{q,k}}$ at each layer $q$ into a single multichannel mapping $\boldcalL_{\boldTheta_q} \colon  \left(\R^J\right)^K \to \left(\R^J\right)^K$  that depends on the entire  binned projection dataset  $\boldy = \{\boldy_k\}$ and on some parameter $\boldTheta_q$. The update from layer $q$ to layer $q+1$ is given by
\begin{equation}\label{eq:unrolling_syn}
	\boldmu^{(q+1)}= \boldcalL_{\boldTheta_{q}} \left(\boldmu^{(q)} \right) \, 
\end{equation}
where the  mapping $\boldcalL_{\boldTheta_q}$ utilizes the entire data and updates the images simultaneously, thus allowing the information to pass between channels. For example, the layer corresponding to Eq.~\eqref{eq:gradient_k} and Eq.~\eqref{eq:smoothing_syn} is $\boldcalL_{\boldTheta_{q}} \left(\boldmu^{(q)} \right)  = \mathbf{prox}^{\boldH}_{\beta_{q} R}\left( \boldmu^{(q)} - \boldH^{-1} \boldg \right)$ with $\boldH = \mathrm{diag}\{\boldH_k\}$ and $\boldg^{(q)} = \left[{\boldg_1^{(q)\top}},\dots, {\boldg_K^{(q)\top}} \right]\transp$. The corresponding  $Q$-layer reconstruction architecture $\boldcalR_{\boldTheta}$, $\boldTheta = \{\boldTheta_q\}$, is given by
\begin{equation}\label{eq:unrolling_all_syn}
	\boldcalR_{\boldTheta^\star}(\boldy) = \boldcalL_{\boldTheta_{Q}^\star} \circ \dots \circ \boldcalL_{\boldTheta_{1}^\star}        \left( \boldmu^{(0)} \right)
\end{equation}
for some initialization $\boldmu^{(0)}$, and the trained parameter $\boldTheta^\star = \{\boldTheta_q^\star \}$ is obtained by supervised training similar to Eq.~\eqref{eq:unrolling_training} but using the data at all energy bins simultaneously:
\begin{equation}\label{eq:unrolling_training_syn}
	\boldTheta^\star = \argmin_{\boldTheta }\,  \bbE \left[  L\left( \boldcalR_{\boldTheta}\left(\boldy^\mathrm{tr}\right)  ,  \boldmu^\mathrm{tr}  \right) \right]\, .
\end{equation}	
A simplified representation of this architecture is shown in Fig.~\ref{fig:unrolling}.    

At the time we are writing this paper, very few research addressed synergistic reconstruction using unrolling-based architectures. We can cite the recent work SOUL-Net by \citeauthor{chen2022soul}~\cite{chen2022soul} which proposes an \ac{ADMM}-based architecture to solve the joint problem~\eqref{eq:synergistic_recon} with the nuclear norm (for \ac{LR} penalty, cf.\@ Section~\ref{sec:syn_recon}) and \ac{TV}. \citeauthor{chen2022soul} modified the singular value thresholding step for nuclear norm minimization by adding a ReLu function with trainable parameters, and replaced the \ac{TV} minimization  with a \ac{CNN} combined with an attention-based network. They showed that their method outperforms ``conventional'' \ac{LR} + sparse decomposition methods.  

Unrolling techniques have shown great promise as a flexible and powerful tool for single-channel image reconstruction. Although these techniques have been applied successfully to a variety of imaging modalities, their application to multichannel synergistic reconstruction in spectral \ac{CT} remains relatively limited and challenging, due to the high-dimensional nature of the data and the need for accurate modeling of the spectral correlations. However, unrolling techniques have been proposed for projection-based and one-step material decomposition, see Section~\ref{sec:mat_decomp}.

\section{Material Decomposition} \label{sec:mat_decomp}
\newcommand\sr[1]{{\color{green}[SR: #1]}}
\newcommand{\AP}[1]{{\color{blue}#1}}
\newcommand{\APrev}[1]{{\color{rev}#1}}

Spectral \ac{CT} techniques such as \ac{DECT} and \ac{PCCT} are often used to characterize the materials of the scanned patient or object by decomposing the linear attenuation coefficient into material images. This process of material decomposition is based on the assumption that the energy dependence of the linear attenuation coefficient in each pixel can be expressed as a linear combination of a small number $M$ of basis functions \cite{alvarez1976energy}. The linear attenuation $\mu(\boldr,E)$ can then be modeled as
\begin{equation}
	\mu (\boldr,E) = \sum_{m=1}^M f_m(E)x_m(\boldr),\label{eq:md_model}
\end{equation}
where $f_m$ represents the $m$\th{} energy-dependent basis function and $x_m$ is the $m$\th{} material image. These basis functions describe physical effects such as photoelectric absorption and Compton scattering \cite{alvarez1976energy} or the linear attenuation coefficients of representative materials of the scanned object such as water and bone for patients. With this model, two basis functions are sufficient to describe the variations of the linear attenuation coefficients of human tissues with energy \cite{bornefalk2012,alvarez2013,tang2021}. One or more basis function(s) may also be used to represent a specific contrast agent, e.g., a material with a K-edge discontinuity in its attenuation coefficient in the range of diagnostic energies (30--140~keV) \cite{roessl2007}. The material images $x_m$ can be represented in the discrete domain as a vector  using the pixel basis functions $u_j(\boldr)$ (see Eq.~\eqref{eq:decomp}) with each pixel of the unknown image decomposed into the chosen material basis. 
The discrete object model for the basis decomposition is then
\begin{equation}\label{eq:mat_decomp}
	\mu(\boldr,E) = \sum_{m=1}^M f_m(E) \sum_{j=1}^J x_{j,m} u_j(\boldr) \, ,  \forall (\boldr,E)\in\R^d\times\R^+ 
\end{equation}
where $x_{j,m}$ is the weight of the $m$\th{} basis function in the $j$\th{} pixel. Injecting \eqref{eq:mat_decomp} into \eqref{eq:beer-lambert_binned} links the material decomposition to the expected value (e.g. the number of detected X-ray photons for \ac{PCCT})
\begin{equation}\label{eq:beer-lambert_md}
	\ybar_{i,k}(\boldx) = \int_{\mathbb{R}^+} h_{i,k}(E) \rme^{ - \sum_{m=1}^M f_m(E) [\boldA_k \boldx_m]_i } \, \rmd E  +   r_{i,k} \, ,
\end{equation}
where $\boldx_m=[x_{1,m},\dots,x_{J,m}]\transp$. Material decomposition aims at estimating the decomposed \ac{CT} images $\boldx=\{ \boldx_m \}$  by matching the expected values $\boldybar(\boldx) = \{ \boldybar_k(\boldx)\}$, $\boldybar_k(\boldx) = [\ybar_{1,k}(\boldx),\dots,\ybar_{I,k}(\boldx)]\transp$, with the measurements $\boldy = \{\boldy_k\}$ with different efficient spectra $h_{i,k}$.

This problem is the combination of two sub-problems: tomographic reconstruction and spectral unmixing. The two problems can be solved sequentially or jointly and most techniques of the literature fall into one of the following categories: image-based, projection-based or one-step material decomposition.

\subsection{Image-based Material Decomposition}\label{sec:im_based_decomp}

Image-based algorithms decompose the multichannel \ac{CT} image $\boldmu = \{\boldmu_k\}$ into material images $\boldx_m$. While each channel $\boldmu_k$ is often obtained by direct methods such as \ac{FBP}, an alternative procedure is the reconstruction of each channel $\boldmu_k$ from $\boldy_k$ by solving the \ac{MBIR} problem in Eq.~\eqref{eq:recon} or the joint reconstruction of $\boldmu = \{\boldmu_k\}$ from $\boldy = \{\boldy_k\}$ by solving the synergistic \ac{MBIR} problem in Eq.~\eqref{eq:synergistic_recon}.
The discretized version of Eq.~\eqref{eq:mat_decomp} is
\begin{equation}\label{eq:mat_decomp_discr}
	\mu_{j,k} = \sum_{m=1}^M F_{k,m} x_{j,m}
\end{equation}
with $F_{k,m}\simeq f_m(E_k)$ and $E_k$ the energy of the attenuation image $\boldmu_k$. The images may be decomposed by solving in each pixel the linear inverse problem
\begin{equation}
	\label{eq:md_image_based}
	\begin{bmatrix}
	  \mu_{j,1}\\
	  \vdots\\
	  \mu_{j,K}
	\end{bmatrix}
	=
	\boldF
	\begin{bmatrix}
	  x_{j,1}\\
	  \vdots\\
	  x_{j,M}
	\end{bmatrix}
\end{equation}
where $\boldF\in\R^{K\times M}$, $[\boldF]_{k,m} = F_{k,m}$, is the same matrix for all voxels characterizing the image-based decomposition problem. It is generally calibrated with spectral \ac{CT} images of objects of known attenuation coefficients. Given that $K$ and $M$ are small, the pseudo-inverse of $\boldF$ can be easily computed and applied quickly after the tomographic reconstruction of $\boldmu$. Image-based material decomposition faces two challenges: (1) the spectral \ac{CT} images are affected by higher noise than conventional \ac{CT} (if the same total dose is split across energy bins) which will be enhanced by the poor conditioning of $\boldF$ and (2) the spectral \ac{CT} images will suffer from beam-hardening artifacts since the efficient spectra $h_{i,k}$ are not truly monochromatic in most cases, i.e., $\boldF$ is actually voxel and object dependent.

Machine learning algorithms have been used for image-based decomposition to mitigate noise and beam-hardening artifacts. Some techniques learn an adequate regularization \cite{wu2020a,wu2021,zhang2023,li2018a,li2020b,li2022b} while using the linear model in Eq.~\eqref{eq:md_image_based}. These techniques are similar in essence to those described in Section~\ref{sec:dict} except that dictionary learning uses decomposed images for spatially regularizing the decomposed images.

\Acp{NN} may be used instead to improve the linear model in Eq.~\eqref{eq:md_image_based} \cite{feng2019}. As in many other fields of research on image processing, deep \acp{CNN} have demonstrated their ability to solve image-based decomposition with a more satisfactory solution than the one produced by a pixel-by-pixel approach.  Several deep learning architectures, previously designed to solve other image processing tasks, have been deployed for image-based decomposition. Most works are based on a supervised learning approach where a dataset of manually segmented basis material images are available: \ac{FCN} \cite{xu2018b}, U-Net \cite{clark2018,abascal2021,holbrook2021,fang2022,fujiwara2022,nadkarni2022}, Butterfly-Net \cite{zhang2019a}, visual geometry group \cite{chen2019,fujiwara2022}, Incept-net \cite{gong2020,gong2022}, \ac{GAN} \cite{shi2021}, Dense-net \cite{wu2019}. These contributions differ on the type of architecture adopted and the complexity of the network which is measured by the number of trainable parameters.
They also differ in which inputs are used by the network, e.g., reconstructed multichannel \ac{CT} images $\boldmu$ \cite{nadkarni2022} or pre-decomposed \ac{CT} images \cite{fang2022}. The network output is generally the decomposed \ac{CT} images $\boldx_m$ but it may also be other images, e.g., the elemental composition \cite{fujiwara2022}, quantities used for radiotherapy planning such as the image of the electron density \cite{su2018a} or the virtual non-calcium image \cite{gong2022}.

\subsection{Projection-based Material Decomposition}\label{sec:proj_based_decomp}

The main limitation of image-based approaches is that the input multichannel \ac{CT} image $\boldmu$ is generally flawed by beam hardening. If several energy measurements are available for the same ray ($\boldA_k=\boldA$ for all $k$), with a dual-layer \ac{DECT} or a \ac{PCCT}, an alternative approach is projection-based decomposition \cite{alvarez1976energy,roessl2007} which aims at estimating projections $a_{i,m}$, $i = 1,\dots,I$, $m=1,\dots,M$, of the decomposed \ac{CT} images $\boldx_m$,
\begin{align}
	a_{i,m} &  = {}  \int_{\calL_i} x_m(\boldr) \,\rmd \boldr   \nonumber \\ 
	        &  = {} [\boldA \boldx_{m}]_{i}, 
\end{align}
from the measurements $\boldy_k$ given the forward model 
\begin{equation}\label{eq:md_projection_based}
	\ybar_{i,k}(\bolda_{i,:}) = \int_{\mathbb{R}^+} h_{i,k}(E) \rme^{ - \sum_{m} f_m(E) a_{i,m} } \, \rmd E  +   r_{i,k}
\end{equation}
where $\bolda_{i,:} = [a_{i,1},\ldots,a_{i,M}]\transp$ and $\bolda_{:,m} = [a_{1,m},...,a_{I,m}]\transp$.  In this context, the expected value $\ybar_k$ becomes a function of $\bolda = \{\bolda_{i,:}\}$ (or $= \{\bolda_{:,m}\}$) instead of $\boldx$. Given the decomposed projections $\bolda_{:,m}$, the images $\boldx_m$ are obtained by solving the following inverse problem 
\begin{equation}\label{eq:dec_proj_inv_prob}
	\boldA \boldx_m = \bolda_{:,m} 
\end{equation}
where multichannel reconstruction algorithm, e.g. those described in Sections~\ref{sec:pen_recon} and \ref{sec:syn_recon} can be deployed to reconstruct $\boldx$ from $\bolda$.

Similar to image-based decomposition, projection-based decomposition can be solved pixel by pixel in the projection domain by solving 
\begin{equation}\label{eq:pixel_proj_recon}
      \hat{\bolda}_{i,:} \in	\argmin_{\bolda_{i,:} \in \R_+^M} \, L\left(\boldy , \boldybar(\bolda_{i,:})\right) + \beta R(\bolda_{i,:}). 
\end{equation}
The number of inputs and unknowns is the same for each projection pixel, but it is more complex because the exponential in Eq.~\eqref{eq:md_projection_based} induces a non-linear relationship between $\ybar_{i,k}$ and $\bolda_{i,:}$. Moreover, this inverse problem \eqref{eq:pixel_proj_recon} is non-convex \cite{abascal2018} (unless, obviously, if the exponential is linearized) and fully-connected \acp{NN} have been used to solve it \cite{lee2012,zimmerman2015}. Such networks can also be used to process input data for spectral distortions before material decomposition \cite{touch2016} or to modify the model described by Eq.~\eqref{eq:md_projection_based} to account for pixel-to-pixel variations \cite{zimmerman2020} or pulse pile-up \cite{jenkins2021}.

However, these approaches cannot reduce noise compared to conventional estimation of most likely solutions \cite{roessl2007} without accounting for spatial variations.
The idea of spatially regularizing pixel-based material decomposition has first been investigated with variational approaches \cite{brendel2016,ducros2017} solving
\begin{equation}\label{eq:proj_recon}
      \hat{\bolda} \in	\argmin_{\bolda \in \left(\R_+^M\right)^I} \, L\left(\boldy , \boldybar(\bolda)\right) + \beta R(\bolda). 
\end{equation}
As in image-based algorithms, \ac{DL} \cite{mechlem2017,mechlem2018a} has been investigated to improve the spatial regularization as well as \acp{CNN} to learn features of the projections with U-Net \cite{abascal2021,holbrook2021}, ResUnet \cite{jiang2023Fully3D}, \ac{SAE} \cite{xu2018a}, perceptron \cite{lu2019}, \ac{GAN} \cite{geng2021} and ensemble learning \cite{lu2015,lu2018}. 

A promising alternative to these supervised techniques, which are learning the physical model from the data, is to solve
\eqref{eq:proj_recon} by combining iterative reconstruction with learning algorithms in so-called learned gradient-descent using unrolling algorithms \cite{eguizabal2022} detailed in Section~\ref{sec:data_learned_syn_recon}. 
Other approaches such as  proposed by \citeauthor{zhang2020}~\cite{zhang2020} combine multiple \acp{NN} both for learning the material decomposition in the projection domain with an additional refinement network in the image domain to enhance the reconstructed image quality.

\subsection{One-step Material Decomposition}\label{sec:onestep_decomp}

One limitation of projection-based decomposition is that some statistical information is lost in decomposed projections $\bolda$ which could be useful to reconstruct the most likely material maps $\boldx$. The noise correlations between the decomposed sinograms $\bolda$ may be accounted for in the subsequent tomographic reconstruction \cite{xu2014a,liu2015d} but it cannot fully characterize the noise of the measurements $\boldy$, in particular with more than two energy bins ($K>2$). Several groups have investigated an alternative solution combining material decomposition and tomographic reconstruction in a one-step algorithm which reconstructs the material maps $\boldx$ from the measurements $\boldy$ by solving the optimization problem
\begin{equation}\label{eq:onestep_recon}
      \boldxhat \in	\argmin_{\boldx \in \left(\R_+^M\right)^J} \, \sum_{k=1}^K L\left(\boldy_k , \boldybar_k(\boldx)\right) + \beta R(\boldx).
\end{equation} 
Compared to Eq.~\eqref{eq:recon}, solving \eqref{eq:onestep_recon} is a far more difficult problem, similar to projection-based algorithms but with a larger number of unknowns ($J\times M$) and inputs ($I\times K$). Several iterative solutions have been proposed to address this problem by optimizing the most likely material maps $\boldx$ given the measurements $\boldy$ with spatial regularization. One of the main differences between these algorithms is the optimization algorithm, from non-linear conjugate gradient \cite{cai2013} to \ac{SQS} algorithms \cite{long2014,weidinger2016,mechlem2018} and primal-dual algorithms \cite{foygelbarber2016,tairi2021}.

The nature of this problem is such that all algorithms based on machine learning have used part of the physical model in their architecture. Generally, combining physics knowledge and deep learning for material decomposition is implemented through unrolling methods \cite{xia2023} (Section~\ref{sec:data_learned_syn_recon}). \citeauthor{eguizabal2022a}~\cite{eguizabal2022a} adapted the projection-based unrolling algorithm of \cite{eguizabal2022}  to one-step reconstruction. The same group has used machine learning to improve the physical model in  Eq.~\eqref{eq:beer-lambert_md} by modeling charge sharing \cite{stroem2022}. Another approach is to insert a backprojection step into the network architecture, i.e. the adjoint of the line integral operator in Eq.~\eqref{eq:beer-lambert_md}, to account for this knowledge in the network architecture \cite{su2022,zhu2022}. Finally, machine learning may be used at each iteration for denoising the images, e.g. with a dictionary approach \cite{zeegers2022}. A self-supervised approach named Noise2Noise prior \cite{fang2021}, which does not require manually segmented ground truth materials images, has been applied to one-step decomposition using a training dataset consisting of sinograms paired with their noisy counterpart obtained by sinogram splitting.

\begin{figure*}
    \centering
    \includegraphics[width=0.9\linewidth]{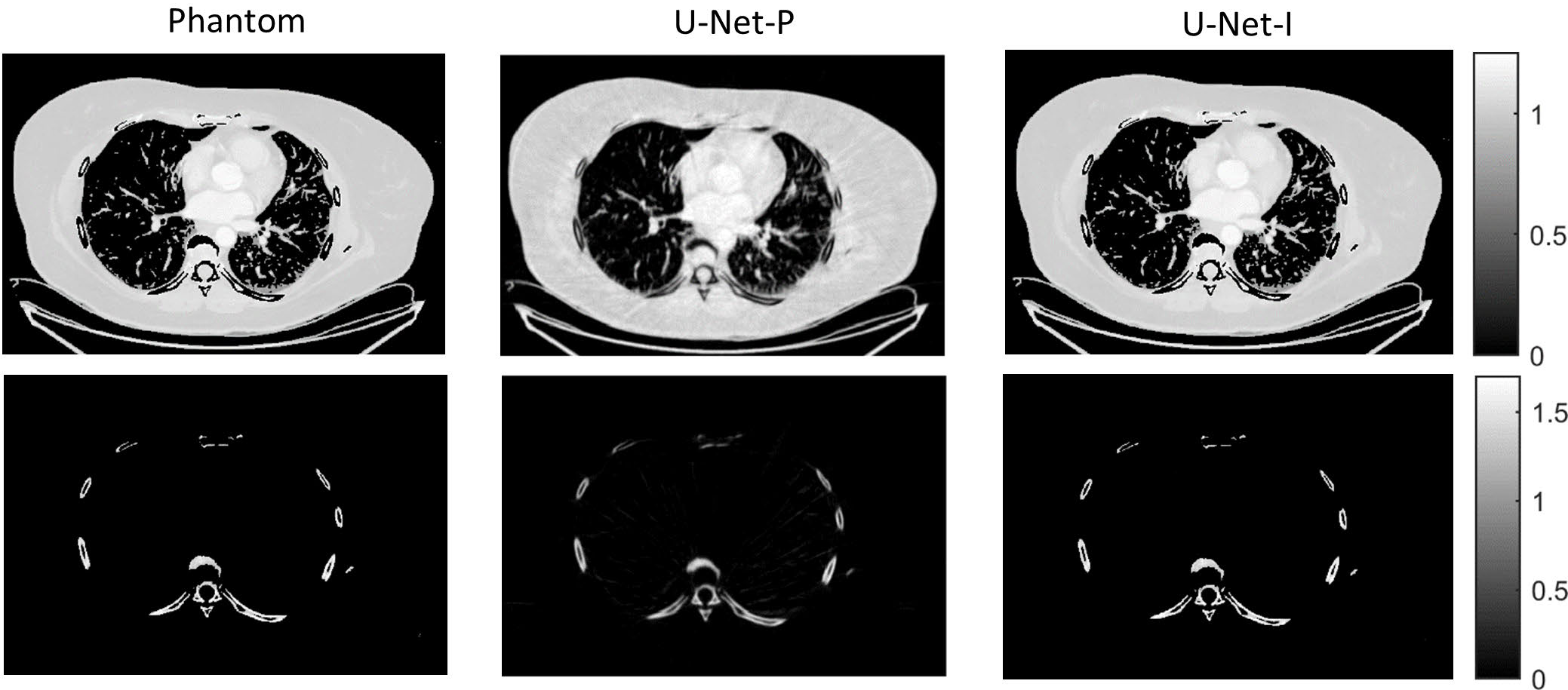}
    \caption{Material decomposition of simulated \ac{PCCT} acquisitions of a patient phantom (left) with projection-based (middle) and image-based (right) U-Net \acp{CNN}. The two materials of the decomposition are soft tissue (top row) and bone (bottom row). Figure adapted from \citeauthor{abascal2021}~\cite{abascal2021} and distributed under a Creative Commons Attribution 4.0 License, see \texttt{https://creativecommons.org/licenses/by/4.0/}.}
    \label{fig:abascal}
\end{figure*}

The different approaches for material decomposition differ on many levels, from computational cost to the accuracy of the decomposed images. For example, \citeauthor{abascal2021}~\cite{abascal2021} compared projection-based and image-based algorithms using variational approaches and machine learning. They observed the best image quality with an image-based material decomposition approach, as illustrated in Fig.~\ref{fig:abascal}. However, the recent Grand Challenge on Deep-Learning spectral Computed Tomography \cite{sidky2023} demonstrated that many different approaches are still under investigation. Nine out of the ten best scorers used machine learning and most combined it with a model of the \ac{DECT} acquisition. The development of such algorithms in clinical scanners will depend on both their practicality, e.g. the computational time, and the accuracy of the material decomposition of real data. 	
\section{Data Pre-processing and Image Post-processing}\label{sec:im_proc}

\Ac{CT} technology has been the front-line imaging tool in emergency rooms due to its fast, non-invasive, and high-resolution features, with millions of scans performed annually worldwide. However, due to the increased cancer incidence from radiation exposure, ``as low as reasonably achievable'' is the central principle to follow in radiology practice. Recent advances in \ac{CT} technology and deep learning techniques have led to great developments in reducing radiation doses in \ac{CT} scans~\cite{vliegenthart2022innovations}. For example, aided by deep learning techniques, much progress has been made in low-dose or few-view \ac{CT} reconstruction without sacrificing significant image quality. Furthermore, the use of \ac{DECT} technology allows further cuts in radiation dose by replacing previous non-contrast \ac{CT} scans with virtual unenhanced images in clinical practice~\cite{virarkar2022virtual}. 

\begin{figure*}
    \centering
    \input{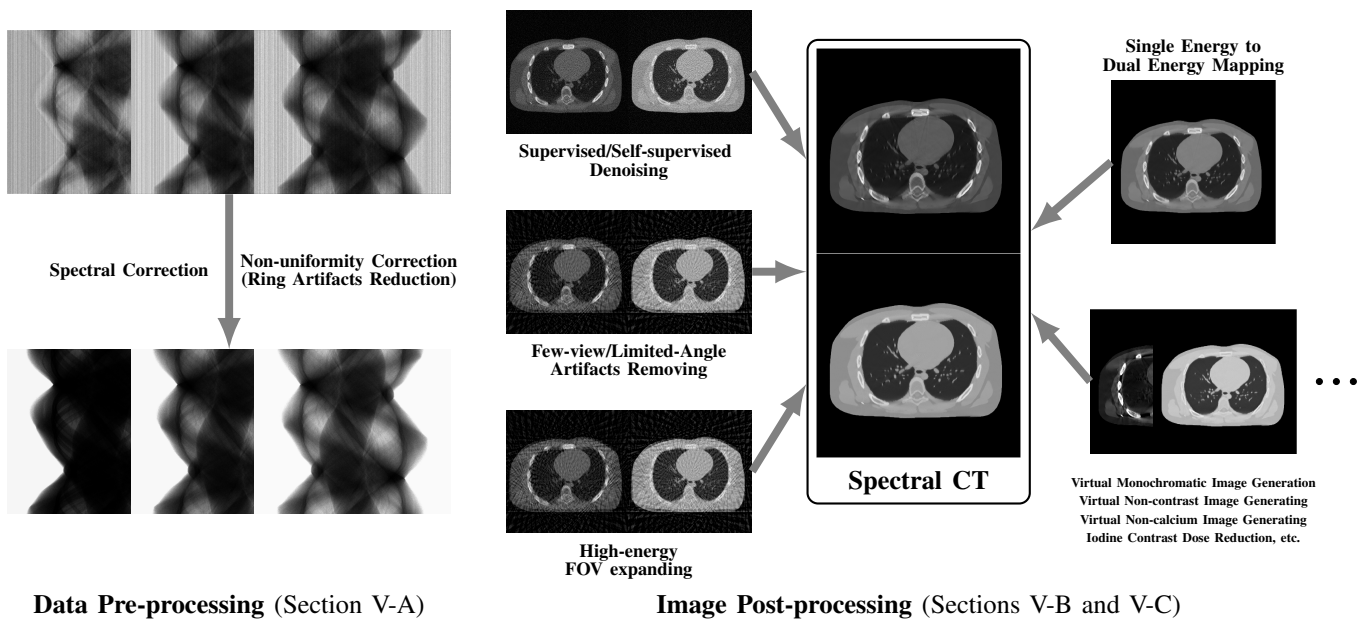}
    \caption{Overview of sub-topics in Section~\ref{sec:im_proc}. The data pre-processing section covers deep correction methods for spectral distortion (e.g., falsely increased counts in the low energy bin due to the charge sharing effect, and non-linear responses due to the pulse pileup effect) and non-uniformity in \ac{PCD} projection images. The image post-processing sections discuss deep post-processing methods to enhance DECT and PCCT imaging and their impacts on clinical diagnosis.}
    \label{fig:Sect5}
\end{figure*}

While many prior-regularized iterative reconstruction techniques described in Section~\ref{sec:recon} inherently suppress noise and artifact, network-based post-processing techniques are also popular for removing noise and artifacts from already reconstructed low-dose spectral images and are covered here. Moreover, \ac{PCCT} with \acp{PCD} is widely viewed as a comprehensive upgrade to \ac{DECT} since it produces less noise, better spectral separation, and higher spatial resolution while requiring less radiation dose~\cite{Willemink2018-oq,Danielsson_2021}. However, the \ac{PCD} often experiences increased nonuniformity and spectral distortion due to charge-sharing and pulse pile-up effects compared to the traditional \ac{EID}, and the correction of these imperfections in \ac{PCD} images is included here. Finally, we also review deep learning techniques that enhance clinical diagnosis with spectral \ac{CT}, which includes virtual monoenergetic image synthesis, virtual noncontrast image generation, iodine dose reduction, virtual calcium suppression, and other applications. The overview of this section is summarized in Fig.~\ref{fig:Sect5}.

\subsection{PCCT Data Pre-processing}\label{sec:data_preproc}

\Acp{PCD} offer much smaller pixel size compared to EIDs and also possess energy discrimination ability that can greatly enhance \ac{CT} imaging with significantly higher spatial and spectral resolution. However, \ac{PCD} measurements are often distorted by undesired charge sharing and pulse pileup effects, which can limit the accuracy of attenuation values and material decomposition. Since accurately modeling these effects is highly complex, deep learning methods are being actively explored for distortion correction in a data driven manner. The initial trial is introduced in \citeauthor{touch2016}~\cite{touch2016} where a simple fully-connected \ac{NN} with two hidden layers of five neurons each  was adopted mainly for charge sharing correction. Later the same network structure but with more neurons was used by~\citeauthor{feng2018neural}~\cite{feng2018neural} to compensate pulse pileup distortion, and similarly in~\cite{fang2020spectral, shi2021x} for spectral distortion correction. A large \ac{CNN} model was first introduced in~\citeauthor{li2020x}~\cite{li2020x} to leverage inter-pixel information for both corrections of charge sharing and pulse pileup effects. The model included a dedicated generator with a pixel-wise fully-connected sub-network for intra-pixel distortion caused by pulse pileup and a convolutional sub-network for inter-pixel cross-talk correction, and was trained using the \ac{WGAN} framework for spectral correction. More recently, \citeauthor{holbrook2021deep}~\cite{holbrook2021deep} used multi-energy \ac{CT} scans with an \ac{EID} to calibrate the \ac{PCD} spectral distortion, and adopted a U-Net to map the distorted \ac{PCD} projections into monochromatic projections generated by multi-energy \ac{CT} scans after material decomposition. \citeauthor{ma2022x}~\cite{ma2022x} introduced CNN-LSTM to correct pulse pileup distortion in X-ray source spectrum measurements, while~\citeauthor{smith2023correcting}~\cite{smith2023correcting} used a spatial-temporal \ac{CNN}  for charge sharing compensation.

There are also several interesting studies on artifact correction for \ac{PCCT} using deep learning methods. \citeauthor{erath2022deep}~\cite{erath2022deep} utilized a U-Net for scatter correction in the sinogram domain to compensate for the Moiré artifacts caused by coarse anti-scatter grids relative to the small detector pixel size, resulting in improved image quality and HU value accuracy. Due to the complexity of \acp{PCD}, their pixels tend to suffer more nonuniformity due to detector imperfections compared to \acp{EID}, making the ring artifact issues more prominent in \ac{PCCT}. To address this issue,
\citeauthor{hein2022deep}~\cite{hein2022deep} trained a U-Net with the perceptual loss for the correction of ring artifacts caused by pixel nonuniformity~\cite{getzin2019non}, while \citeauthor{fang2020removing}~\cite{fang2020removing} used two U-Nets in both projection domain and image domain for ring artifacts removal.

\subsection{Image Post-processing}\label{sec:im_postproc}

\subsubsection{Image Denoising}\label{sec:im_denoising}

In \ac{CT} imaging, it is important to limit the radiation dose to patients, but reducing the dose often gives rise to image noise, which can strain radiologists' interpretation. To address this issue, various image denoising methods have been developed that aim to recover a clean version $\boldmu^\star$ from a noisy image $\boldmu^0$ by leveraging prior knowledge $R$ of the image to maintain sufficient image quality for clinical evaluation,
\begin{equation}
    \boldmu^\star = \argmin_{\boldmu}\left\|\boldmu- \boldmu^0\right\|_2^2 + \beta R(\boldmu).
    \label{eq:denoisingProblem}
\end{equation}
The development of \ac{CT} noise reduction techniques has a long history with its root dating back shortly after the invention of \ac{CT}. While our focus is on deep learning and spectral \ac{CT}, it is important to briefly cover classic post-processing denoising techniques and deep learning techniques for single energy \ac{CT}, as they can still be applied to spectral \ac{CT} in a channel-by-channel manner. We will then dive into recent trends of self-supervised learning deep denoising methods, as well as deep methods that incorporate the correlations between energy channels.

Spatial filtering methods leverage the statistical nature of noise fluctuations and are achieved through local averaging or nonlocal averaging means \cite{kachelriess2001generalized,manduca2009projection,li2014adaptive}; optimization-based denoising methods, on the other hand, incorporate image model preassumptions such as domain sparsity, piecewise linearity, or gradient smoothness as regularization. Some well-known methods in this category include \ac{TV}~\cite{tian2011low}, \ac{DL}~\cite{mairal2009online,xu2012low}, wavelet based denoising~\cite{jiao2008multiscale}, \ac{BM3D}~\cite{makinen2020collaborative}, and others. A good discussion of these classic denoising techniques is provided by~\citeauthor{diwakar2018review} in their review paper~\cite{diwakar2018review}. Different from the explicitly defined prior knowledge in traditional methods, the development of deep learning techniques, particularly \acp{CNN}, provides a data-driven approach to learn the implicit distribution knowledge from large amounts of images, offering a one-step solution to the denoising problem (Eq.~\eqref{eq:denoisingProblem}), i.e.,
\begin{equation}
    \boldmu^\star = \boldcalF_{\boldtheta^\star}\left(\boldmu^0\right), \label{eq:inferencing}
\end{equation}
where $\boldcalF_{\boldtheta^\star}$ denotes the network function with optimized parameters $\boldtheta^\star$ after training. Since they are way more powerful than the traditional methods, deep methods will soon dominate the research field of \ac{CT} image denoising. Initially, these methods were primarily trained in a supervised fashion using paired noisy and clean images, as generally depicted by Eq.~\eqref{eq:supervisedLearning}, and the successful examples include REDCNN~\cite{chen2017low}, wavelet network~\cite{kang2017deep} and stacked competitive network~\cite{du2017stacked}.
\begin{equation}
    \boldtheta^\star = \argmin_{\boldtheta}\bbE\left[\mathcal{L}\left(\boldcalF_{\boldtheta}\left(\boldmu^0\right),\boldmu^1\right)\right],
    \label{eq:supervisedLearning}
\end{equation}
where $\mathcal{L}(\cdot,\cdot)$ denotes a general loss function for network training, $\bm{\mu}^1$ is the  clean image corresponding to the noisy one $\bm{\mu}^0$ and the expectation is taken over pairs $(\bm{\mu}^0,\bm{\mu}^1)$ from the training dataset. Following the idea, various network structures and loss functions have been explored. Representative network structures include U-Net~\cite{shan20183,gunduzalp20213d,liu2018low}, DenseNet~\cite{ming2020low}, \ac{GAN}~\cite{yang2018low,wolterink2017generative,tang2019unpaired,you2019ct}, ResNet~\cite{yang2017improving,yin2019domain}, Residual dense network~\cite{nishii2022deep}, Quadratic neural network~\cite{fan2019quadratic}, transformer~\cite{wang2021ted}, diffusion model~\cite{xia2022low}, and more. Commonly used loss functions include \ac{MSE}, \ac{MAE}, structural similarity index~\cite{wang2004image,zhang2020metainv,unal2022unsupervised}, adversarial loss~\cite{goodfellow2014generative, yang2018low}, \ac{TV} loss~\cite{feng2020preliminary,chen2021low}, perceptual loss~\cite{ledig2017photo, yang2018low}, edge incoherence~\cite{shan2019competitive}, identity loss~\cite{zhu2017unpaired,kang2019cycle,you2019ct}, projection loss~\cite{unal2022unsupervised}, and more. For more detailed information, we refer readers to the latest two review papers on low-dose \ac{CT} denoising~\cite{kulathilake2021review, immonen2022use}.

The issue of missing paired labels was soon realized when researchers attempted to apply supervised methods in practice. To address this, a number of unsupervised or self-supervised methods have been proposed. For instance, cycle-\ac{GAN} based techniques are able to utilize unpaired data for training by promoting cycle consistency between domains~\cite{you2019ct,kang2019cycle,park2019unpaired,tang2019unpaired}. However, these GAN-based methods have been criticized for potentially generating erroneous structures. Poisson Unbiased Risk Estimator (PURE) and Weighted Stein's Unbiased Risk Estimator (WSURE) are alternative methods that convert the supervised \ac{MSE} loss calculation into a form that only relies on the noisy input, the network output, and its divergence~\cite{kim2020unsupervised}. This approach forms an unsupervised training framework where the divergence term is approximated using Monte-Carlo perturbation method~\cite{lehtinen2018noise2noise}. Noise2Noise is another method that enables us to train the network with paired noise-noise images which are equivalent to being trained with original noise-clean pairs,
\begin{equation}
    \boldtheta^\star = \argmin_{\boldtheta}\bbE\left[\left\|\boldcalF_{\boldtheta}\left(\boldmu^0\right) - \boldmu^1\right\|_2^2\right],
    \label{eq:unsupervisedLearning}
\end{equation}
where $\boldmu^0$ and $\boldmu^1$ are different noisy realizations of the same image, e.g., two independent \ac{CT} scans of the same object. Building on this idea, several recent variant methods have been developed for self-supervised low-dose \ac{CT} denoising by generating noisy pairs via various approaches~\cite{hasan2020hybrid,won2021self,yuan2020half2half,zhang2022s2ms,zhang2021noise2context,wu2021low,hendriksen2020noise2inverse,niu2022noise,unal2021self}. For instance, Noise2Inverse proposes to partition projection data into several sets and enforcing consistency between corresponding reconstructions~\cite{hendriksen2020noise2inverse}, while Noise2Context promotes similarity between adjacent \ac{CT} slices in \ac{3D} thin-layer \ac{CT}~\cite{zhang2021noise2context}; Half2Half adopts the thinning technique~\cite{elhamiasl2019simulating} to split a full dose real \ac{CT} scan into two pseudo half dose scans~\cite{yuan2020half2half}.

Spectral \ac{CT} powerfully extends the conventional single energy \ac{CT} by introducing an extra energy dimension. However, the splitting of photons into different energy bins increases the noise level of the projection at each bin compared to conventional \ac{CT} with the same overall radiation dose. Therefore, to achieve optimal denoising performance for spectral \ac{CT}, it is necessary to leverage inter-bin information, similar to the approach taken in learned synergistic reconstruction (Section~\ref{sec:data_learned_syn_recon}), as described below,
\begin{equation}
    [\boldmu_1^\star,\dots,\boldmu_K^\star] = \boldcalF_{\boldtheta^\star}\left(\boldmu_1^0,\dots,\boldmu_K^0\right)
    \label{eq:multichannel} \, .
\end{equation}
Several recent papers have explored this direction. ULTRA~\cite{wu2021deep} incorporates an $\ell^p$-norm and anisotropic total variation loss to train a residual U-Net with multichannel inputs from \ac{PCCT} scans. Noise2Sim~\cite{niu2022noise} constructs noisy pairs using the Noise2Noise principle and replaces each pixel from the original noisy image with one of its $k$-nearest pixels searched from the spatial dimension (including adjacent slices) and measured by non-local means. The multichannel image is fed to the network as a whole, and its value from different bins can be constructed independently to fully leverage the self-similarities within the spectral \ac{CT} scans. By this means, comparable or even better performance has been demonstrated on experimental \ac{PCCT} scans against the supervised learning methods. S2MS~\cite{zhang2022s2ms} proposes another interesting approach to leverage the inter-channel information by converting the linear attenuation map from each channel to a channel-independent density map, which forms different noisy realizations of the density images from multiple channels. Promising results from this self-supervised learning idea are demonstrated on a simulation study.

Besides developing various deep denoising methods, researchers have also investigated the effects of noise reduction on the downstream tasks~\cite{evans2022effects,wu2021deep}. For example, \citeauthor{evans2022effects}~\cite{evans2022effects} compared the material decomposition results of multi-bin \ac{PCCT} images before and after denoising with \ac{BM3D} and Noise2Sim through phantom studies. They found that image denoising improves the accuracy of material concentration quantification results, but not material classification results. In the clinical domain, there are several \ac{FDA}-approved deep denoising methods from multiple vendors (e.g., the TrueFidelity from GE Healthcare, the Advanced Intelligent Clear-IQ Engine (AiCE) from Canon, PixelShine from Algomedica, ClariCT.AI from ClariPI Inc., etc), and numerous studies have been performed to investigate their impacts on clinical significance. For ease of notation, we use \acf{DLIR} to refer specially to these \ac{FDA}-approved methods in clinical applications. \citeauthor{noda2021low}~\cite{noda2021low} showed that with \ac{DLIR}, the radiation dose of whole-body \ac{CT} can be reduced by up to 75\% while maintaining similar image quality and lesion detection rate compared to standard-dose \ac{CT} reconstruction with iterative reconstruction through a study cohort of 59 patients. This conclusion is also supported in other studies where \ac{DLIR} and iterative reconstruction of the same patient scans are compared, showing that \ac{DLIR} provides significantly preferred image quality and reduced noise~\cite{parakh2021sinogram,fair2022image}.

For the diagnosis with \ac{DECT}, the pancreatic cancer diagnostic acceptability and conspicuity can be significantly improved, and the use of \ac{DLIR} reduces the variation in iodine concentration values while maintaining their accuracy~\cite{noda2022deep}. \citeauthor{fukutomi2023deep}~\cite{fukutomi2023deep} suggests similar results in terms of iodine concentration quantification through both phantom and clinical studies. The stability of iodine quantification accuracy with \ac{DLIR} has also been investigated in the context of radiation dose variation. For example, \citeauthor{kojima2021novel}~\cite{kojima2021novel} found that the accuracy is not affected by the radiation dose when the dose index is greater than 12.3~mGy. For a more detailed assessment of \ac{DLIR} in clinical practice, a recent review paper by~\citeauthor{szczykutowicz2022review}~\cite{szczykutowicz2022review} provides a good starting point. It is also worth noting that the aforementioned studies with \ac{PCCT}~\cite{evans2022effects} and \ac{DECT}~\cite{fukutomi2023deep} lead to different conclusions about the impacts of denoising on iodine/material concentration quantification, which could be attributed to the different energy discrimination mechanisms between \ac{PCCT} and \ac{DECT}, as the number of energy bins and spectral separation can significantly influence the accuracy and stability of material decomposition performance~\cite{Willemink2018-oq}.

\subsubsection{Artifacts Correction}

Besides noise, image artifact is another factor that affects the quality of \ac{CT} image for diagnostic evaluation. Few-view or limited-angle reconstruction is an effective method to reduce the radiation dose, but it can introduce globally distributed artifacts that are difficult to remove. To be concise and avoid overlap with Section III, here we only cover recent progress on post-processing-based artifact reduction approaches via deep learning for spectral \ac{CT}. The networks are often trained in a supervised manner for this application and directly applied to \ac{FBP} reconstructions to remove artifacts, which can be similarly described as Eq.~\eqref{eq:supervisedLearning} and Eq.~\eqref{eq:inferencing} with $\boldmu^0$ and $\boldmu^1$ being few-view/limited-angle reconstruction and full-view/full-angle reconstruction respectively. For example, to reduce few-view reconstruction artifacts and accelerate reconstruction for scans at multiple energy points (i.e., 32~channels), \citeauthor{mustafa2020sparse}~\cite{mustafa2020sparse} proposed a U-Net-based approach that maps few-view \ac{FBP} reconstruction images to computationally intensive full-view iterative reconstruction images with \ac{TV} regularization. The 32-channel \ac{FBP} images were fed to the network simultaneously and transformed to high-quality 32-channel reconstructions in one step, majorly reducing the computational cost. More recently, \citeauthor{lee2021ultra}~\cite{lee2021ultra} developed a multi-level wavelet convolutional neural network, using a U-Net architecture with the wavelet transform as the down-sampling/up-sampling operations, that effectively captures and removes globally distributed few-view artifacts. The network simultaneously processes multi-channel images to leverage inter-channel information, and demonstrates promising results both numerically and experimentally with an edged silicon strip \ac{PCD}. To address limited-angle artifacts for cone beam \ac{DECT}, \citeauthor{zhang2023time}~\cite{zhang2023time} proposed the TIME-Net, which utilizes a transformer module with global attention. In addition, the two complementary limited-angle scans at two energies are fused together to form a prior reconstruction, then the features extracted from the prior reconstruction, high-energy reconstruction, and low-energy reconstruction are fused in latent space to leverage inter-channel information with the network.

In dual-source \ac{DECT} scanners, the high-energy imaging chain (i.e., tube B with a tin filter, typically at 140~keV) often has a restricted \ac{FOV} (e.g., 33cm) due to physical constraints compared to the other chain (e.g., 50cm for tube A), which can be problematic for larger patients and affect diagnosis. To outpaint the missing regions and match the size of normal \ac{FOV}, \citeauthor{liu2021unsupervised}~\cite{liu2021unsupervised} proposed a self-supervised method that maps the low-energy image to the high-energy image with a loss function only focusing on image values within the restricted \ac{FOV}. The outpainting is then automatically completed leveraging the shift-invariant nature of \acp{CNN}. Similarly, \citeauthor{schwartz2021evaluating}~\cite{schwartz2021evaluating} proposed a method for \ac{FOV} extension that involves feeding both the high-energy image and the low-energy image in the network, along with a high-energy estimation from the low-energy image via a piecewise-linear transfer function. The trained network was applied to patient data for renal lesion evaluation and showed reliable results in terms of HU value and lesion classification accuracy in the extended regions.

\subsection{Image Generation for Clinical Applications}\label{sec:im_synthesis}

With the recent development of \ac{DECT} and \ac{PCCT} techniques, spectral imaging is reshaping the clinical utilization of \ac{CT}. These techniques enable the generation of multiple types of images that enhance diagnosis and improve disease management, such as \acp{VMI}, virtual unenhanced images, bone suppression images, and material decomposition maps. A good number of research studies have been performed in these areas using deep learning approaches.

\subsubsection{Single-Energy to Dual-energy Mapping}\label{sec:SE2DE}

Despite the great possibilities offered by \ac{DECT} and \ac{PCCT}, their accessibility remains limited in comparison to conventional single-energy \ac{CT}, largely due to the high cost involved. To bridge the gap, \citeauthor{zhao2019deep}~\cite{zhao2019deep} successfully demonstrated the feasibility of using deep learning to predict high-energy \ac{CT} images from given low-energy \ac{CT} images in a retrospective study. Shortly, \citeauthor{lyu2021estimating}~\cite{lyu2021estimating} proposed a material decomposition \ac{CNN} capable of generating high-quality \ac{DECT} images from a low-energy scan combined with a single view high-energy projection, leveraging the anatomical consistency and energy-domain correlation between two energy images in \ac{DECT}. The feasibility of this method has been validated with patient studies, showing great potential for simplifying \ac{DECT} hardware and reducing radiation exposure during \ac{DECT} scans.

\subsubsection{Virtual Monochromatic Image}
\acp{VMI} are widely used as the basis for routine diagnosis due to their ability to reduce beam-hardening and metal artifacts, and enhance iodine conspicuity. They are obtained by linearly combining the basis material volume fraction maps~\cite{alvarez1976energy,yu2011virtual} obtained after material decomposition, as described by the  material decomposition model in Section~\ref{sec:mat_decomp}. To enhance readability and clarity, Eq.~\eqref{eq:md_model}, which outlines this model, is replicated here in a spatially discrete form:
\begin{equation}
    \boldmu(E) = \sum_{m=1}^M{f_m(E)\boldx_m},
\end{equation}
where $\boldx_m$ denotes the volume fraction map of the $m$\th{} material basis, $f_m(E)$ stands for the linear attenuation coefficient of the corresponding material at energy $E$, and $M$ is the total number of material basis. However, the synthesis of \acp{VMI} relies on material decomposition results and is therefore limited to \ac{DECT} and \ac{PCCT}, which may not be available in less developed areas. Similar to section~\ref{sec:SE2DE}, a number of approaches have been explored to directly synthesize the \acp{VMI} from single-energy \ac{CT} scans. \citeauthor{cong2020virtual}~\cite{cong2020virtual} first used a modified ResNet for \ac{VMI} generation from single polychromatic \ac{CT} scans, then developed a sinogram domain method~\cite{cong2021monochromatic} synthesizing \acp{VMI} with a fully-connected \ac{NN} for virtual monochromatic energy sinogram prediction from single polychromatic measurements. \citeauthor{kawahara2021image}~\cite{kawahara2021image} employed a \ac{GAN} to generate \acp{VMI} from equivalent keV-\ac{CT} images, while~\citeauthor{koike2022pseudo}~\cite{koike2022pseudo} used a U-Net for a similar purpose in imaging of head and neck cancers. More interestingly, \citeauthor{fink2022jointly}~\cite{fink2022jointly} found that using \acp{VMI} synthesized from single-energy \ac{CT} images for pulmonary embolism classification provides better performance compared to working directly on the original single-energy images.

On the other hand, \ac{VMI} synthesis is a downstream task after image reconstruction and material decomposition, during which deep denoising plays a role and potentially affects \ac{VMI} quality in clinical practice. Extensive studies have investigated this effect through quantitative assessment and/or subjective reader studies. \citeauthor{kojima2021novel}~\cite{kojima2021novel} examined \ac{VMI} \ac{CT} number accuracy at various radiation doses, finding that accuracy remains unaffected except at extremely low radiation doses (6.3~mGy). \citeauthor{sato2022deep}~\cite{sato2022deep} compared \acp{VMI} from \ac{DLIR} with routine baselines from hybrid iterative reconstruction for contrast-enhanced abdomeninal \ac{DECT} imaging, concluding that vessel and lesion conspicuity of \acp{VMI} and iodine density images are improved with \ac{DLIR}. \citeauthor{xu2022quantitative}~\cite{xu2022quantitative} reached a similar conclusion, and particularly they found that 40~keV \acp{VMI} from \ac{DLIR} poses better \ac{CNR} and similar or improved image quality compared to 50~keV \ac{VMI} from hybrid iterative reconstruction, suggesting that 40~keV \ac{VMI} with \ac{DLIR} could be a new standard for routine low-keV \ac{VMI} reconstruction. The study for carotid \ac{DECT} angiography by~\citeauthor{jiang2022deep}~\cite{jiang2022deep} also supports the conclusion that \ac{DLIR} improves the image quality and diagnostic performance of \acp{VMI} compared to hybrid iterative reconstruction. This superiority is further confirmed in \ac{DECT} angiography with reduced iodine dose (200~mgI/kg) in terms of image quality and arterial depiction by~\citeauthor{noda2022deep}~\cite{noda2022deep}.
Additionally, the effect of direct denoising on \acp{VMI} has been investigated. In a study of~\citeauthor{lee2022deep}~\cite{lee2022deep} the post-processed \ac{VMI} using ClariCT.AI (a \ac{FDA}-approved vendor-agnostic imaging denoising software) is compared with original standard \ac{VMI} in the assessment of hypoenhancing hepatic metastasis. The results suggest denoising leads to better image quality and lesion detectability. A similar conclusion was achieved by~\citeauthor{seo2022deep}~\cite{seo2022deep} with the same post-denoising method for the evaluation of hypervascular liver lesions.

\subsubsection{Contrast Agent Dose Reduction}

Iodine-enhanced \ac{CT} is essential for diagnosing various diseases. However, iodine-based contrast media can cause significant side effects, including allergic reactions in certain patients, and dose-dependent kidney injury and thyroid dysfunction. To investigate the possibility of reducing iodine administration dose while maintaining diagnostic accuracy, \citeauthor{haubold2021contrast}~\cite{haubold2021contrast} trained a GAN to selectively enhance iodine contrast. They ultimately achieved a 50\% contrast dose saving ratio, confirmed by a visual Turing test involving three radiologists assessing pathological consistency. \citeauthor{noda2022radiation}~\cite{noda2022radiation} explored the potential of leveraging vendor \ac{DLIR} for simultaneous iodine and radiation dose reduction in thoraco-abdomino-pelvic \ac{DECT} imaging.
They compared the 40~keV \acp{VMI} from \ac{DLIR} of double low-dose (50\% iodine, 50\% radiation) scans with \acp{VMI} from the hybrid iterative reconstruction of standard dose scans. The diagnostic image quality was achieved in 95\% of participants in the double low-dose group, suggesting the feasibility of maintaining diagnostic quality at half doses of radiation and iodine using \ac{DLIR}.

\subsubsection{Others}

Several other intriguing deep post-processing techniques for spectral \ac{CT} include virtual non-contrast image synthesis, virtual non-calcium image synthesis, and spectral \ac{CT}-based thermometry. Virtual non-contrast images can replace non-contrast scans in a \ac{DECT} scanning protocol, thus saving radiation dose. However, pure physics-based two-measurement material decomposition algorithms exhibit limited accuracy and stability in the presence of three materials. \citeauthor{poirot2019physics}~\cite{poirot2019physics} employed a \ac{CNN} to leverage the anatomic information, bridging the gap between material decomposition-derived virtual non-contrast images and the real non-contrast images to generate higher fidelity images. 

Virtual non-calcium images are useful for visualizing bone marrow, osteolytic lesions, and even the diagnosis of multiple myeloma~\cite{kosmala2018multiple,kosmala2018dual}. Like virtual non-contrast images, they also suffer from excessive noise and artifacts resulting from material decomposition. \citeauthor{gong2022}~\cite{gong2022} proposed a custom dual-task \ac{CNN} that directly maps the input of spectral \ac{CT} images to material type maps and corresponding mass density maps. The experimental results demonstrate significantly reduced noise and artifacts in virtual non-calcium images and great visibility of bone marrow lesions. 

\Ac{CT}-based thermometry provides a non-invasive method for estimating temperature inside the human body by monitoring the attenuation value changes associated with temperature-dependent radiodensity. \citeauthor{heinrich2022ct}~\cite{heinrich2022ct} explored the potential of improving temperature sensitivity with \acp{VMI} from \ac{DLIR} of \ac{DECT} scans compared to conventional single-energy \ac{CT} images. Their results show that \acp{VMI}  significantly enhances temperature sensitivity for different materials, particularly for bone with a boost of 211\%. The application of \ac{DLIR} and hybrid iterative reconstruction has no effect on temperature measurement, suggesting the great potential for dose reduction with deep learning techniques. More recently, \citeauthor{wang2023photon}~\cite{wang2023photon} incorporated an advanced \ac{PCD} with 4 energy bin measurements for robust material decomposition and a fully-connected \ac{NN} for temperature prediction. They observed a non-linear relationship between thermal sensitivity and the concentration of CaCl$_2$ solution in the experiment, achieving final thermometry accuracies of $3.97^\circ$C and $1.8^\circ$C for 300~mmol/L CaCl$_2$ solution and a milk-based protein shake, respectively.

\section{Perspectives}\label{sec:perspectives}

Advances in spectral \ac{CT} is a major frontier of the medical \ac{CT} field, which combines cutting-edge hardware for photon-counting detection and \ac{AI}-empowered software for deep learning-based reconstruction. As we have reviewed above, photon-counting spectral \ac{CT} promises to significantly improve the medical \ac{CT} performance  in terms of spatial resolution, spectral resolution, tissue contrast, and dose efficiency. The distinguished capability of photon-counting \ac{CT} in material decomposition is clinically attractive to perform novel multi-contrast-enhanced studies and boost \ac{CT}, not only in anatomical imaging but also functional or even cellular imaging tasks. All of these can be implemented using machine learning methods or coupled with machine learning methods. Most of such machine learning methods are deep neural networks, involving each key step in the whole imaging workflow.

Looking ahead, the convergence of photon-counting and deep-learning techniques will surely establish spectral \ac{CT} as the new standard of medical \ac{CT}. To realize the huge potential of photon-counting spectral \ac{CT}, there remain challenges to be addressed before task-specific methods and protocols can be successfully translated into clinical practice. These challenges include but are not limited to the following aspects.

\paragraph*{Direct Reconstruction}

Deep \acp{NN} have been explored to reconstruct images from sinograms in a number of studies. In this  approach, a neural network is trained on a large set of sinogram-image pairs until the network predicts realistic reconstructed images. Here, the \ac{NN} learns to reconstruct the image and at the same time to reduce noise and to incorporate any other corrections desirable for reconstruction. Early methods developed for tomographic reconstruction using deep networks include AUTOMAP \cite{zhu2018image} for \ac{MR} reconstruction as well as 
LEARN \cite{chen2018learn} and iCT \cite{li2019learning}  for \ac{CT} reconstruction. To tackle the computational complexity, more sophisticated and efficient networks were developed \cite{thaler2018sparse,xie2019deep,jiao2021dual,kandarpa2022lrr}. 

Direct reconstruction techniques may be extended to multichannel reconstruction including photon-counting spectral \ac{CT} reconstruction. One possible way would be to have multi-channel networks incorporating data in multiple energy bins or an ensemble of networks with weight sharing for each energy. Importantly, correlations among these data in these channels should be utilized; for example, as a term in the loss function.

\paragraph*{Locally linear embedding Motion Correction}

The much-reduced pixel size of \acp{PCD} enables \ac{CT} imaging at ultra-high resolution, which is one major advantage of \acp{PCCT} over traditional \ac{EID}-based CT and critical to resolve anatomical and pathologic details, such as cochlear features, lung nodules, and coronary plaques. As resolution drastically improves, the sensitivity to patient motion and geometric misalignment becomes high and can be the limiting factor of image resolution. This increased sensitivity also challenges the assumption of smooth patient movement across views~\cite{jang2019head,sun2016iterative,sisniega2017motion}.

\begin{figure}
	\center
	\includegraphics[width = 0.9\linewidth]{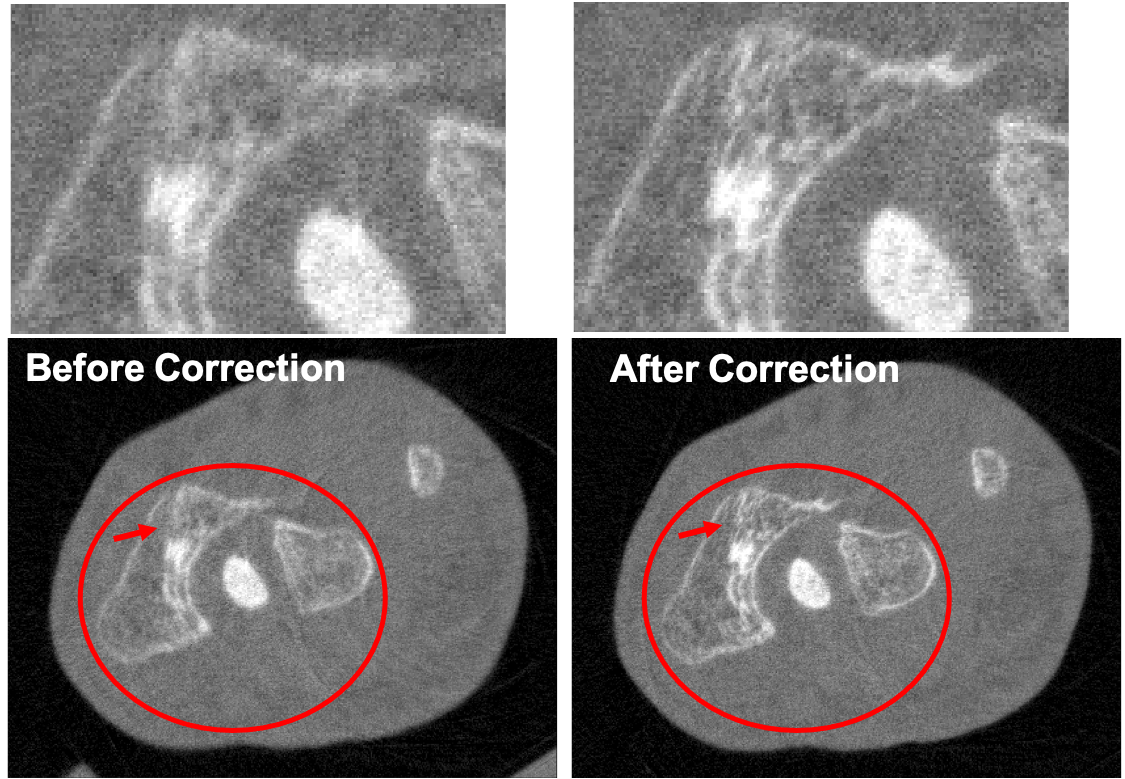}
	\caption{High-resolution \ac{PCCT} scan of a patient wrist from a clinical trial (90 $\mu$m voxel) before and after motion correction (Adapted from \citeauthor{li2022motion}~\cite{li2022motion} 
		with permission).}\label{fig:LLEhelical}
\end{figure}

To address the issue, \citeauthor{li2022motion}~\cite{li2022motion} developed a rigid patient motion compensation method for high-resolution helical \ac{PCCT} based on \ac{LLE}. Their method is in a coarse-to-fine searching framework to boost efficiency, along with several accuracy improving steps masking bad pixel, unreliable volume and patient bed respectively. The method was evaluated on patient wrist scans in a clinical trial, revealing fine bony structures previously hidden by motion blur, as shown in Fig.~\ref{fig:LLEhelical}. Subsequently, \citeauthor{li2022motion2}~\cite{li2022motion2} proposed a unified reference-free all-in-one motion correction method for robotic \ac{CT} with arbitrary scanning trajectories using a nine-degree-of-freedom model, which is capable of addressing rigid patient motion, system misalignment, and coordination errors simultaneously.
The effectiveness of the method has been verified on experimental robotic-arm-based \ac{PCCT} scans of a sacrificed mouse demonstrating a great resolution boost and artifacts reduction.

\paragraph*{Diffusion Models}

As a score-matching-based generative approach, the \acp{DM} have recently drawn a major attention of the community as they effectively compete or even outperform \acp{GAN} for image generation and other tasks \cite{dhariwal2021diffusion}, and have been broadly adapted for medical imaging \cite{kazerouni2023diffusion}, including \ac{PCCT} image generation \cite{hein2023generation}. They involve gradually degrading a sample of interest (i.e., an image) with subtle Gaussian noise until the sample becomes a random Gaussian field, learning the noising process in terms of a score function, and then, by inversion from a Gaussian noise realization,  generate a meaningful sample~\cite{ho2020denoising}. Specifically, the inverse process uses the gradient of the log-density of the prior (the score) which is approximated with a \ac{NN} trained for score matching, and generates an image according to the a-priori probability distribution of the training dataset.

\Acp{DM} can be used to solve inverse problems by adding a data fidelity gradient descent step in the inverse diffusion, or by using the pseudo  a-posteriori probability distribution conditioned to the observed data, which work in an unsupervised manner. These methods have been used in various inverse problems such as deblurring on RGB multichannel images \cite{chung2022diffusion}. Moreover, the \acp{DM} are independent of the measurement model, and the same approaches can be used in multi-energy spectral \ac{CT} reconstruction or one-step material decomposition under different imaging geometries and sampling conditions. Recently,

\paragraph*{Hardware Refinement} Over the past years, photon-counting detectors have been greatly refined. There are more efforts on CZT detectors, but deep-silicon detectors are also of great interest. While CZT detectors and alike are more compact, the silicon technology is more mature, reliable and cost-effective with the potential to give more quantitative spectral imaging results. A detailed comparison is yet to be seen. Since the photon-counting detector pitches are substantially smaller than that of the energy-integrating detectors, the spatial resolution of \ac{CT} images can be accordingly improved, coupled with a reduced X-ray source focal spot. However, a small focal spot usually means a low X-ray flux. Hence, the balance must be made between image resolution, noise and imaging speed. It is underlined that while the hardware refinement in either detectors or sources is important, this kind of research will be more often performed by leading companies than academic groups. Since this review is more focused on computational aspects of spectral \ac{CT}, in the following we discuss more \ac{AI}-related challenges.
	
\paragraph*{Big Data Construction} It is well known that big data is a prerequisite for data-driven research. Clearly, it is not easy to have big \ac{PCCT} data for several reasons, including limited accessibility to \ac{PCCT} scans, patient privacy, industrial confidentiality, and so on. We believe that this issue must be addressed using simulation tools, and ideally done in a healthcare metaverse. Such an idea was discussed as the first use case in a recent perspective article \cite{wang2022development}. Along that direction, virtual twins of physical \ac{PCCT} scanner models can scan patient avatars to produce simulated data. Along a complementary direction, a limited number of real \ac{PCCT} scans can be used to train a generative model for realistic image augmentation.  For example, it was recently shown that the diffusion model can be used to synthesize realistic data with suppressed privacy leakage \cite{shi2023conversion}. This will facilitate federated learning at the level of datasets.

\paragraph*{AI model Development}
	
 When sufficiently informative \ac{PCCT} data are available, more advanced \ac{AI} models should be developed to address current weaknesses of deep reconstruction networks in the \ac{CT} field. The well-known problems of deep networks include stability, generalizability, uncertainty, interpretability, fairness, and more. As briefly mentioned in our review, a unique opportunity in deep learning-based \ac{PCCT} imaging is raw data correction for charge-sharing, pile-up and other effects. These effects are very complicated, nonlinear and stochastic, but deep learning-based solutions are few and there will be more in the future. Furthermore, large models are gaining great attention, with ChatGPT as a precursor of the next generation of \ac{AI} methods, i.e., as the first step into the future of \acf{AGI}. It is believed that large models, multi-modal large models in particular, will further improve the \ac{PCCT} performance.
	
\paragraph*{High-performance and High-efficiency Computing} 
	
Deep learning with large models takes computational resources. Parallel/cloud computing, model distillation and hybrid (combination of classic and deep learning) reconstruction methods can be synergistic to develop practical \ac{PCCT} methods. Special hardware such as FPGAs \cite{cong2011high} could be adapted in \ac{PCCT} tasks for imaging speed and energy efficiency.

\paragraph*{Clinical Translation}	

The development of accurate and robust \ac{PCCT} methods should lead to diverse clinical applications, from screening and diagnosis to treatment planning and prognosis. \ac{PCCT} can be also used to guide minimally invasive procedures, such as biopsy and ablation, by providing real-time information over a region of interest \cite{li2020clinical}. The integration of \ac{PCCT} (and \ac{DECT}) with other imaging modalities, such as \ac{MRI} and \ac{PET}, would be beneficial as well, leading to a better understanding of anatomical forms and pathological functions.

\paragraph*{Hybrid PET/CT Spectral Imaging}

The integration of spectral CT with PET has the potential to open novel clinical applications. However, such an integrated system either requires a costly hardware upgrade or is associated with increased radiation exposure. Most existing spectral CT imaging methods are based on a single modality that uses X-rays. Alternatively, it is possible to explore a combination of X-ray and $\gamma$-ray for spectral imaging \cite{Wang20pmb}. The concept of this \ac{PET}-enabled spectral CT method exploits a standard time-of-flight \ac{PET} emission scan to derive high-energy $\gamma$-ray \ac{CT} attenuation images and combines the images with low-energy X-ray CT images to form dual-energy or multi-energy imaging. This method has the potential to make spectral CT imaging more readily available on clinical PET/CT scanners. The enabling algorithm of this hybrid spectral imaging method is the reconstruction of $\gamma$-ray attenuation images from PET emission data using the maximum-likelihood attenuation and activity algorithm \cite{Rezaei12tmi, Wang20pmb}. While the counting statistics of PET emission data are relatively low, machine learning-based approaches have been developed to further improve image reconstruction, for example, using the kernel method alone \cite{Wang20pmb, Wang15tmi} or in combination with deep neural networks \cite{Li21ptrsa, Li21spie, Li22tmi}. These reconstruction approaches are directly based on single subjects without requiring pretraining from a large number of datasets. Alternatively, many other big data-based deep learning techniques that are described in Section \ref{sec:recon}, Section \ref{sec:mat_decomp}, and Section \ref{sec:im_proc} may be applied to the development of hybrid PET/CT spectral imaging.

\section{Conclusion}

In conclusion, this review has systematically reviewed spectral \ac{CT} with an emphasis on photon-counting and deep learning techniques. This field has evolved from traditional \ac{DECT} with an established status in medical imaging to contemporary \ac{PCCT} with promising results and new utilities. Several remaining challenges have been discussed. The future of this technology looks exciting, with numerous opportunities for us to explore so that our imaging dreams can be turned into reality.
	
\section*{Acknowledgment}
All authors declare that they have no known conflicts of interest in terms of competing financial interests or personal relationships that could have an influence or are relevant to the work reported in this paper. 

\AtNextBibliography{\footnotesize} 

\printbibliography

\end{document}